\documentclass[10pt,conference,final]{IEEEtran}
\IEEEoverridecommandlockouts
\usepackage{cite}
\usepackage{amsmath,amssymb,amsfonts}
\usepackage{algorithmic}
\usepackage{graphicx}
\usepackage{textcomp}
\usepackage{xcolor}
\usepackage[normalem]{ulem}

\usepackage[utf8]{inputenc}
\usepackage{etoolbox}
\usepackage{units}
\usepackage{url}

\usepackage{hyperref}

\usepackage[switch]{lineno}

\usepackage[frozencache,cachedir=.]{minted}
\input{default.pygstyle}

\makeatletter
\AtBeginEnvironment{minted}{\dontdofcolorbox}
\def\dontdofcolorbox{\renewcommand\fcolorbox[4][]{##4}}
\makeatother

\newtoggle{releaseStuffAfterDoubleBlind}
\toggletrue{releaseStuffAfterDoubleBlind}

\newtoggle{draftComments}
\togglefalse{draftComments}

\usepackage{xargs}
\iftoggle{draftComments}{%
  \usepackage[textsize=scriptsize,textwidth=1.4in]{todonotes}
  \addtolength{\marginparwidth}{1.2in}
  \addtolength{\marginparsep}{0.2in}
  \addtolength{\hoffset}{1.2in}
  \addtolength{\paperwidth}{2.4in}
  \newcommand{\extras}[1]{{\color{blue}{#1}}}%
  \pagestyle{plain}
}{%
  \usepackage[textsize=scriptsize,textwidth=1.4in,disable]{todonotes}%
  \newcommand{\extras}[1]{}%
}

\iftoggle{releaseStuffAfterDoubleBlind}{
\newcommand\changed[1]{#1}
}{
\newcommand\changed[1]{\textcolor{blue}{#1}}
}
\newcommand\change[1]{\changed{#1}}

\begin{document}

\title{High-Performance GPU-to-CPU Transpilation and Optimization via High-Level Parallel Constructs}

\iftoggle{releaseStuffAfterDoubleBlind}{%
    \author{%
        \IEEEauthorblockN{%
            \parbox{\linewidth}{\centering
                William S. Moses\IEEEauthorrefmark{1},
                Ivan R. Ivanov\IEEEauthorrefmark{2},
                Jens Domke\IEEEauthorrefmark{3},
                Toshio Endo\IEEEauthorrefmark{2},
                % Charles Leiserson\IEEEauthorrefmark{1}%
                Johannes Doerfert\IEEEauthorrefmark{5}, and
                Oleksandr Zinenko\IEEEauthorrefmark{4}
            }%
        }%
        \vspace{1ex}
        \IEEEauthorblockA{%
            \small{%
                \IEEEauthorrefmark{1}
                Massachusetts Institute of Technology, USA
                \texttt{~~wmoses@mit.edu}%
            }%
        }%
        \IEEEauthorblockA{%
            \small{%
                \IEEEauthorrefmark{2}
                Tokyo Institute of Technology, Japan
                \texttt{~~ivanov.i.aa@m.titech.ac.jp, endo@is.titech.ac.jp}%
            }%
        }%
        \IEEEauthorblockA{%
            \small{%
                \IEEEauthorrefmark{3}
                RIKEN Center for Computational Science, Japan
                \texttt{~~jens.domke@riken.jp}%
            }%
        }%
        \IEEEauthorblockA{%
            \small{%
                \IEEEauthorrefmark{5}
                Argonne National Laboratory, USA
                \texttt{~~jdoerfert@anl.gov}%
            }%
        }%
        \IEEEauthorblockA{%
            \small{%
                \IEEEauthorrefmark{4}
                Google, France
                \texttt{~~zinenko@google.com}%
            }%
        }%
    }%
}%

\maketitle

\begin{abstract}

While parallelism remains the main source of performance, architectural implementations and programming models change with each new hardware generation, often leading to costly application re-engineering. Most tools for performance portability require manual and costly application porting to yet another programming model.

We propose an alternative approach that automatically translates programs written in one programming model (CUDA), into another (CPU threads) based on Polygeist/MLIR. Our approach includes a representation of parallel constructs 
that allows conventional compiler transformations to apply transparently and without modification and enables parallelism-specific optimizations. 
We evaluate our framework by transpiling and optimizing the CUDA Rodinia benchmark suite for a multi-core CPU and achieve a 76\% geomean speedup over handwritten OpenMP code.
Further, we show how CUDA kernels from PyTorch can efficiently run and scale on the CPU-only Supercomputer Fugaku without user intervention. \changed{Our PyTorch compatibility layer making use of transpiled} CUDA PyTorch kernels outperforms the PyTorch CPU native backend by \changed{2.7$\times$}.

\end{abstract}

\section{Introduction}

The end of single-core performance scaling means that parallelism and domain-specificity are now the main sources of efficiency increases.
Supercomputer architects compete in ingenuity to support compute- and memory-intensive applications from physics simulations to machine learning.
The latest and fastest supercomputer, Fugaku, is based exclusively on A64FX CPUs that, unlike commodity CPUs, provide support for high-bandwidth memory access and energy efficiency comparable to that of recent GPUs~\cite{fugaku}.

However, efficient and productive use of such computers for practical applications is challenging as recent frameworks and high-performance libraries have been developed with NVidia GPUs in mind. 
For example, attempts to port PyTorch~\cite{pytorch} to A64FX have met multiple challenges. 
The ``native'' default CPU PyTorch backend provides only na\"ive versions for critical kernels,
such as 2D convolution implemented as six nested loops.
Intel’s oneDNN~\cite{oneDNN} unsurprisingly performs poorly for Arm CPUs since it is tailored for commodity CPUs without high-bandwidth memory.
Fujitsu's customized oneDNN~\cite{dnnl_aarch64} is better tuned, but not universally competitive with GPUs.

\begin{figure}[htbp]
\begin{minted}[fontsize=\footnotesize, escapeinside=||]{cuda}
__device__ float sum(float* data, int n) { ... }
__global__
void normalize(float *out, float* in, int n) {
  int tid = blockIdx.x + blockDim.x * threadIdx.x;
  // Optimization: Compute the sum once per block.
  // __shared__ int val;
  // if (threadIdx.x == 0) val = sum(in, n);
  // __syncthreads;
  float val = sum(in, n);
  if (tid < n)
    out[tid] = in[tid] / val;
}
void launch(int *d_out, int* d_in, int n) {
  normalize<<<(n+31)/32, 32>>>(d_out, d_in, n);
}
\end{minted}
\vspace*{-3mm}
\caption{A sample CUDA program \texttt{normalize}, which normalizes a vector and the CPU function \texttt{launch} which calls the kernel. Presently, the call to \texttt{sum} is called in each thread, leading to a total of $O(N^2)$ work. The work can be partially reduced to $O(N^2/B)$ through the use of shared memory (in the comments above), which enables the sum to be computed once per block, or completely reduced to $O(N)$ by computing \texttt{sum} once before the kernel.}
\label{fig:intro_example}
\end{figure}

Many non-library approaches to achieve performance portability have been put forth. They range from language extensions such as OpenCL~\cite{opencl} or OpenACC~\cite{openacc} to parallel programming frameworks such as Kokkos~\cite{kokkos}, to domain-specific languages such as \textsc{Spiral}~\cite{spiral}, Halide~\cite{halide} or Tensor Comprehensions~\cite{tc}. All these approaches still require \change{legacy} applications to be \change{ported, and sometimes entirely rewritten,} due to differences in the language, or the underlying programming model\changed{, of the original program and the target framework}.

We explore an alternative approach based on a fully automated compiler that takes code in one programming model (CUDA) and produces a binary targeting another one (CPU threads). While GPU-to-CPU translation has been explored in the past~\cite{mcuda,diamos2010ocelot,han2021cox}, it was rarely able to produce efficient code.
In fact, optimizations for CPUs and even generic compiler transforms, such as common sub-expression elimination or loop-invariant code motion, are hindered by the lack of analyzable representations of parallel constructs inside the compiler~\cite{moses2017should}.
As an example, consider the summation in Figure~\ref{fig:intro_example} which could be done by a single thread per block, or by a single thread in total once all computation happens on the CPU.
As representations of parallelism within a mainstream compiler have only recently begun to be explored
in a mainstream compiler~\cite{schardl2019tapir, kotsifakou2018hpvm, stelle2017openmpir, DBLP:conf/lcpc/DoerfertF18, DBLP:conf/iwomp/DoerfertDF19}, existing transformations are limited and tend to apply to simple CPU codes only.

We propose a compiler model for most common GPU constructs: multi-level parallelism, level-wide synchronization, and level-local memory.
\changed{This differs from CPU parallelism, which provides a single level of parallelism, a unified memory and peer synchronization. In contrast to source and AST-level approaches, which operate before the optimization pipeline, and existing compiler approaches, which model synchronization as a ``black-box'' optimization barrier, we model synchronization entirely from memory semantics. This both allows synchronization-based code to inter-operate with existing optimizations and enables novel parallel-specific optimizations.} % In contrast to other approaches for GPU-to-CPU transpilation or heterogeneous libraries which operate on the source or AST level, our framework operates fundamental 

Our model is implemented in the MLIR layer~\cite{mlir} of the LLVM compiler infrastructure~\cite{llvm} and it leverages MLIR's nested-module approach for GPU codes~\cite{mlir_gpu}
We extended the Polygeist~\cite{polygeist} C/C++ frontend to support CUDA and to produce MLIR which preserves high-level parallelism and program structure.
Our prototype compiler is capable of compiling PyTorch CUDA kernels, as well as other compute-intensive benchmarks, to any CPU architecture supported by LLVM.
In addition to transformations accounting for the differences in the execution model, we also exploit parallelism on the CPU via OpenMP.
Finally, our MocCUDA PyTorch integration uses our approach to compile and execute CUDA kernels in absence of a GPU while substituting unsupported calls.

The correctness and efficiency of our end-to-end translation is evaluated by compiling Rodinia CUDA benchmarks~\cite{rodinia} as well as PyTorch CUDA kernels.
When targeting a commodity CPU, our OpenMP-accelerated CUDA code yields comparable performance with the reference OpenMP implementations from the Rodinia suite, as well as improved scalability. When using our framework to run PyTorch on the CPU-only Fugaku Supercomputer, we achieve roughly twice the images processed per second by the \texttt{conv2d} kernel from Resnet-50~\cite{he2016deep} compared to the OneDNN-based PyTorch CPU backend, and comparable performance to the hand-tuned overall training.

Overall, our paper makes the following contributions:
\begin{itemize}
    \item \changed{A common high-level and platform-agnostic representation of SIMT-style parallelism backed by a semantic definition of barrier synchronization that ensures correctness through memory semantics, which ensures transparent application of existing optimizations.}
    \item Novel parallel-specific optimizations which can exploit our high-level parallel semantics to optimize programs.
    \item An extension to the Polygeist C/C++ frontend for MLIR which is capable of directly mapping GPU and CPU parallel constructs into our high-level parallelism primitives.
    \item An end-to-end transpilation of CUDA to CPU for a subset of the Rodinia~\cite{rodinia} benchmark suite and the internal CUDA kernels within PyTorch~\cite{pytorch} necessary to run a Resnet-50 on the CPU-only Fugaku supercomputer.
\end{itemize}

\section{Background}
\label{sec:background}

Mainstream compilers like Clang and GCC lack a unified high-level representation of parallelism. 
Specifically, compiling parallel constructs in frameworks like CUDA, OpenMP, or SYCL, forces the body of a parallel region to exist within a separate (closure) function which is invoked by the respective runtime.
Concepts such as thread index or synchronization are then represented separately, often through opaque intrinsic calls.
As the compiler historically lacked information about parallelism and effects of the involved runtimes, any parallel construct also inadvertently acted as a barrier to optimization. 
While there have been attempts~\cite{DBLP:conf/lcpc/DoerfertF18,DBLP:conf/iwomp/DoerfertDF19,DBLP:conf/sc/TianSSLGCMSRDZR17,moses2017should, schardl2019tapir, kotsifakou2018hpvm, stelle2017openmpir} in recent years to improve representations for CPU parallelism constructs, accelerator programming comes with additional challenges.
The unique programming model and complex memory hierarchy have caused high-level representations of GPU parallelism within mainstream compiler remain under-explored.

\begin{figure}[tbp]
\begin{tabular}{@{}l}
\begin{minipage}[T]{\linewidth}
\begin{minted}[fontsize=\footnotesize, escapeinside=||]{llvm}
target triple = "x86_64-unknown-linux-gnu"

define @launch(float* %d_out, float* %d_in, i32 %n) {
  call @__cudaPushCallConfiguration(...)
  call @__cudaLaunchKernel(@normalize_stub, ...)
  ret
}
\end{minted}
\vspace{0.3em}
\end{minipage}
\\\hline
\begin{minipage}{\linewidth}
\vspace{0.3em}
\begin{minted}[fontsize=\footnotesize, escapeinside=||]{llvm}
target triple = "nvptx64"

define @normalize(float* %out, float* %in, i32 %n) {
  %tid = call i32 @llvm.nvvm.ptx.tid.x()
  %sum = call i32 @sum(i32* %in, i32 %n)
  %cmp = icmp slt i32 %tid, %n
  br i1 %cmp, label %body, label %exit
body:
  %gep = getelementptr float* %in, i32 %tid
  %load = load float, float* %gep
  %nrm = fdiv float %load, %sum
  %ptr = getelementptr float* %out, i32 %tid
  store float %nrm, float* %ptr
  br label %exit
exit:
  ret
}
\end{minted}
\end{minipage}
\end{tabular}
\vspace*{-1mm}
    \caption{Simplified lowering of the \texttt{launch} and \texttt{normalize} functions from Figure \ref{fig:intro_example}, when compiled with LLVM/Clang.
    As the two functions emit different assembly codes, they are placed in separate modules with no context for how, or if, they are called.
   % Moreover, the GPU kernel function \texttt{normalize} does not appear to have its body run multiple times, let alone is denoted that it is run in parallel.
    }
    \label{fig:llvm_example}
\end{figure}

\subsection{GPU Compilation}

Consider the CUDA program in Figure \ref{fig:intro_example},
which normalizes a vector. 
When compiled using LLVM/Clang, the GPU program is a separate compilation unit, as shown in Figure \ref{fig:llvm_example}
This prevents any optimization between the GPU kernel and the CPU calling code.
In the case of Figure \ref{fig:intro_example}, the total work of the program in a traditional compiler is $O(N^2)$, due to the $O(N)$ call to \texttt{sum} being performed for each thread.
However, if the call to \texttt{sum} could be performed only once prior to the kernel call, e.g., by performing the loop-invariant code motion (LICM) transformation, the work would reduce to $O(N)$. 
A less effective variant of this optimization could reduce the work to $O(\frac{N^2}{B})$ through the use of shared memory.
MLIR provides a nested-module representation for GPU programs that supports host/device code motion~\cite{mlir_gpu}, but parallel code motion has not been implemented.
In GPU to CPU code motion, LICM out of a parallel loop is always legal as formerly device memory would also be available on the host.

\begin{figure}[tbp]
\begin{minted}[fontsize=\footnotesize, escapeinside=||]{mlir_lexer.py:MlirLexer -x}
// Kernel launch is available within the calling
// function, enabling optimizations across the
// GPU/CPU boundary.
func @launch(%h_out : memref<?xf32>,
             %h_in  : memref<?xf32>, %n : i64) {
  // Parallel for across all blocks in a grid.
  parallel.for (%gx, %gy, %gz) = (0, 0, 0) 
            to (grid.x, grid.y, grid.z) {
  
    // Shared memory = stack allocation in a block.
    %shared_val = memref.alloca : memref<f32>
    
    // Parallel for across all threads in a block.
    parallel.for (%tx, %ty, %tz) = (0, 0, 0) 
              to (blk.x, blk.y, blk.z) {
      // Control-flow is directly preserved. 
      if %tx == 0 {
        %sum = func.call @sum(%d_in, %n)
        memref.store %sum, %shared_val[] : memref<f32>
      }
      // Syncronization via explicit operation.
      polygeist.barrier(%tx, %ty, %tz)
      %tid = %gx + grid.x * %tx
      if %tid < %n {
        %res = ...
        store %res, %d_out[%tid] : memref<?xf32>
      }
    }
  }
}
\end{minted}
\caption{Representation of the shared-memory variant of the CUDA \texttt{launch}/\texttt{normalize} code from Figure \ref{fig:intro_example} in Polygeist/MLIR. The kernel call is made available directly in the host code which calls it. The parallelism is made explicit through the use of parallel for loops across the blocks and threads, and shared memory is placed within the block to signify it can be accessed from any thread in the same block, but not from a different block. }
\label{fig:par_example}
\end{figure}

\subsection{MLIR Infrastructure}
MLIR is a recent compiler infrastructure designed for reuse and extensibility~\cite{mlir}.
Its internal representation (IR) can be seen as a conceptual successor on the widely adopted LLVM IR~\cite{llvm} with better support for programming models beyond the conventional CPU one.
Rather than providing a predefined set of instructions and types, MLIR operates on collections of \emph{dialects} that contain sets of interoperable user-defined operations, attributes and types.
Operations are a generalization of IR instructions that can be arbitrarily complex, in particular, contain regions with more IR thus creating a nested representation.
Operations define and use values that obey single static assignment (SSA)~\cite{ssa}.
For example, MLIR dialects may model entire physical or virtual instruction sets such as NVVM (virtual IR for Nvidia GPUs)%
, other IRs such as LLVM IR, higher-level control flow constructs such as affine loops, parallel programming models such as OpenMP and OpenACC, machine learning graphs, etc.
MLIR supports and encourages to mix operations from different dialects in the same compilation unit.

MLIR supports GPU thanks to the eponymous dialect, which defines the high-level SIMT programming model as well as host/device communication mechanisms, and a set of low-level platform-specific dialects: NVVM (CUDA), ROCDL (ROCm) and SPIR-V.
The particularity of MLIR's approach to GPU programming consists in its \emph{unified} code representation.
Thanks to the flexibility of the IR, a module may contain other modules, e.g., the ``host'' translation unit may embed the ``device'' translation unit as IR rather than file reference or binary blob.
This approach provides host/device optimization opportunities unavailable to other compilers, in particular by freely moving code between host and device~\cite{mlir_gpu}.

\subsection{Polygeist}
Polygeist is a C and C++ frontend for MLIR based on Clang~\cite{polygeist}.
It is capable of translating a broad range of C++ programs into a mix of MLIR dialects that preserve elements of the high-level structure of the program.
In particular, Polygeist preserves structured control flow (loops and conditionals) as MLIR SCF dialect.
It also simplifies analyses by preserving multi-dimensional array constructs whenever possible by relying on the MLIR's multi-dimensional memory reference (memref) type.
Finally, Poylgeist is able to identify parts of the program suitable for polyhedral optimization~\cite{polyhedral} and represent them using the Affine dialect.

\section{Approach}
\label{sec:approach}

We extended the Polygeist compiler~\cite{polygeist} to directly emit parallel MLIR from CUDA.
This leverages the unified CPU/GPU representation to allow the optimizer to understand host/device execution, and to enable optimization across kernel boundary.
The use of existing MLIR's first-class parallel constructs (\texttt{scf.parallel}, \texttt{affine.parallel}) enables us to target existing CPU and GPU backends.
Finally, MLIR's extensible operation set allows us to define custom instructions, with relevant properties and custom optimizations.

We define the representation of a GPU kernel launch as follows (illustrated in Fig.~\ref{fig:par_example}):
\begin{itemize}
    \item A 3D parallel for-loop over all blocks in the grid.
    \item A stack allocation for any shared memory, scoped to be unique per block.
    \item A 3D parallel for-loop over all threads in a block.
    \item A custom Polygeist barrier operation that provides equivalent semantics to a CUDA/ROCm synchronization.
\end{itemize}

This procedure enables us to represent any GPU program in a form that preserves the desired semantics. It is fully understood by the compiler and is thus amenable to compiler optimization.
Moreover, by representing GPU programs with general parallelism, allocation, and synchronization constructs, we are not only able to optimize the original program, but also retarget it for a different architecture.

\subsection{Barrier Semantics}
\label{sec:barrier_transform}
A CUDA or ROCm \texttt{\_\_syncthreads} function guarantees that all threads in a block have finished executing all instructions prior to the function call, before any threads executes any instruction after the call. Traditionally, compilers represent such functions as opaque optimization barriers that could touch all memory, and forbid any transformation involving them.

\begin{figure}
\begin{minted}[fontsize=\footnotesize, escapeinside=||]{cuda}
__global__ f() {
  codeA();
  barrier();
  codeB();
}
\end{minted}
\caption{A program containing a barrier between two arbitrary instructions.}
\label{fig:barrier}
\end{figure}

In our system, we chose to represent such thread-level synchronization through a new \texttt{polygeist.barrier} operation. Unlike other approaches, \texttt{polygeist.barrier}  (hence referred to as simply \texttt{barrier}) aims to only prevent transformations that would change externally visible behavior. Rather than disallowing any code motion across a \texttt{barrier}, we can successfully achieve the desired semantics by defining \texttt{barrier} to have specific memory properties\changed{, represented as a collection of memory locations (including unknown), and memory effect type (read, write, allocate, free), as is standard within MLIR.}
Consider the simple program in Fig.~\ref{fig:barrier}.
The impact of the synchronization can only be observed if \texttt{codeA} and \texttt{codeB} access the same memory. Moreover, if both only read the same memory location, the synchronization is also unnecessary. We can then enumerate the three remaining cases:
\begin{enumerate}
    \item \texttt{codeA} writes, \texttt{codeB} loads
    \item \texttt{codeA} loads, \texttt{codeB} writes
    \item \texttt{codeA} writes, \texttt{codeB} writes
\end{enumerate}

The barrier having the write behavior of \texttt{codeA} would ensure correctness of case 1: the load in \texttt{codeB} could not be hoisted above the barrier, as it would appear to read a different value. 
Symmetrically, the barrier having the write behavior of \texttt{codeB} would ensure correctness of case 2.
Thus, the union of the writing behaviors of \texttt{codeA} and \texttt{codeB} is sufficient to prevent loads from being incorrectly moved across the barrier.

However, this does not prevent writes from being moved.
For example, \texttt{codeB} could be duplicated above the barrier in case 2, and it would appear to have the same final memory state since the extraneous write before the barrier would never be read. Thus, we also define the barrier to have the reading behavior of \texttt{codeA} and \texttt{codeB}.

\change{
This can be extended to include memory effects of all operations in the parallel loop which may have been executed before, or after, a given barrier. On a control flow graph with explicit branches, this can be done by exploring the operations within predecessors or successors, respectively. However, operating on MLIR's structured control flow level, with explicit operations for loops and conditionals, allows the analysis to be simplified. Furthermore, if more than one barrier is present in the same block, it is unnecessary to look past it.
}

Given a sufficiently expressive side effect model, the memory semantics of the barrier can be further expanded.
Barriers order reads/writes to the same location from \emph{different} threads, the natural execution order is sufficient within one thread.
Therefore, barriers need not capture the memory effects of operations where the address is an \emph{injective function} of the thread identifier.
Raising memory accesses into linear forms available in the Affine dialect, when possible, enables precise analysis.
Consider the code in Fig.~\ref{fig:barrier_semantics_holes}.
The read and write expressions around the barrier have the affine access sets $\mathcal{A}_o = \{A(i): i = tx\}$ where $tx$ is the thread \texttt{x} identifier.
The barrier has the affine access set $\mathcal{A}_b = \{A(i): i \neq tx \}$.
Since the sets of accessed addresses do not overlap, $\mathcal{A}_o \cap \mathcal{A}_b = \varnothing$, code motion across the barrier is allowed.
Conceptually, the write to \texttt{A[threadIdx.x]} always happens before the read within the same thread so the barrier is unnecessary. \changed{In contrast, if the load and store to \texttt{A} were offset by 1, the barrier would be necessary as the data loaded after the barrier would be stored by a different thread.} \changed{More generally, when defining the memory properties of a given barrier by collecting the memory locations and effect types from all relevant operations, one only needs to include memory locations used by a different thread.}
Aliasing guarantees must be checked when more than one base address is involved.

\begin{figure}
\begin{minted}[fontsize=\footnotesize]{cuda}
__global__ f() {    // 0 <= t.x < blockDim.x
  A[threadIdx.x] = ...; //  W A[i]: i == t.x
  barrier();            // RW A[i]: i != t.x
  ... = A[threadIdx.x]; // R  A[i]: i == t.x
}
\end{minted}
\caption{Barrier semantics can be refined to accessing memory addresses accessed by operations above/below it in all threads \emph{except} the current one.}
\label{fig:barrier_semantics_holes}
\end{figure}

\subsection{Barrier Lowering}
\label{sec:barrier_lowering}
To enable GPU programs to run on a CPU, we must efficiently emulate the synchronization behavior of GPU programs. Whereas the memory semantics in Section~\ref{sec:barrier_transform} enable us to preserve the correctness of barriers during optimization, this section discusses how to implement the barrier on a CPU. 

CPU architectures have no notion of thread blocks, nor the barrier instruction which waits on this conceptual grouping of threads.
Instead, we use regular CPU threads and work sharing to distribute the thread-block loop iterations across them.
Conceptually, this is different from the GPU execution model in which numerous explicit threads execute one iteration each.
Work sharing requires each thread to execute multiple iterations sequentially, making it impossible to synchronize in the middle of iterations, but only at the end of the loop.

To address this, we developed a new barrier elimination technique for our MLIR representation.
As discussed in Section~\ref{sec:related_work}, several approaches have been explored in the past including loop fission and continuation passing. 
Our approach is an extension of the former combining two styles of transformations: \emph{parallel loop splitting} and \emph{parallel loop interchange}.

\subsubsection{Parallel Loop Splitting}
\label{sec:loopsplitting}
\begin{figure}
    \centering

\begin{tabular}{@{}l|c@{}}
\begin{minipage}[T]{0.45\linewidth}
\begin{minted}[fontsize=\footnotesize, escapeinside=||]{mlir_lexer.py:MlirLexer -x}


parallel %i = 0 to 10 {
  %x = load data[%i]
  %y = load data[2 * %i]
  %a = fmul %x, %x
  %b = fmul %y, %y
  %c = fsub %x, y
  barrier
  call @use(%a, %b, %c)
  ...
}

| |
\end{minted}
\end{minipage}&%
\begin{minipage}[T]{0.48\linewidth}
\begin{minted}[fontsize=\footnotesize, escapeinside=||]{mlir_lexer.py:MlirLexer -x}
%x_cache = memref<10xf32>
%y_cache = memref<10xf32>
parallel %i = 0 to 10 {
  %x = load data[%i]
  %y = load data[2 * %i]
  store %x, %x_cache[%i]
  store %y, %y_cache[%i]
}
parallel %i = 0 to 10 {
  %x = load %x_cache[%i]
  %y = load %y_cache[%i]
  %a = fmul %x, %y
  %b = fsub %y, %z
  call @use(%a, %b)
  ...
}
\end{minted}
\end{minipage}
\end{tabular}
    \caption{Example of a parallel loop splitting around a barrier. Here the code above the barrier (comprising 2 loads and 3 floating point operations) is placed in a separate parallel for loop from the code following the barrier (comprising a call to \texttt{@use} and other operations). This transformation eliminates the barrier, while preserving the semantics. As the values \texttt{\%a}, \texttt{\%b}, and \texttt{\%c} are used after the barrier, extra care must be taken to ensure they are available. Here, the min-cut algorithm preserves \texttt{\%x} and \texttt{\%y}, and then recompute \texttt{\%a}, \texttt{\%b}, and \texttt{\%c} as this would result in 2 values being preserved rather than 3.}
    \label{fig:parallel_split}
\end{figure}
Suppose a barrier has the kernel function (or, in our representation, parallel \texttt{for} loop) as its direct parent. It can be eliminated by splitting the loop around the barrier into two parallel \texttt{for} loops that run the code before and after the barrier, respectively. If the code before the barrier created SSA values that were used after it, these must be either stored or recomputed in the second parallel loop. We use the technique similar to one in~\cite{moses2021reverse} to determine the minimum amount of data that needs to be stored. Specifically, we can represent the problem by creating a graph of all SSA values. We then mark each value definition that cannot be recomputed (e.g. loads from overwritten memory) before the \texttt{barrier} as source, and values used after the \texttt{barrier} as sinks. By performing a minimum branch cut on this graph, we can derive the minimum amount of data that needs to be stored.

\subsubsection{Parallel Loop Interchange}
Not all barrier operations have a parallel \texttt{for} as their immediate parent, some may be nested in other control flow operations.
We created a model that specifies what instructions may run in parallel.
With the sole exception of \texttt{barrier}, our representation does not require any specific ordering or concurrency to the program.
Therefore it is legal (though potentially a reduction in parallelism) to add additional barriers. We can use this property to implement barrier lowering for control flow.

Consider a control-flow construct \texttt{C} containing a barrier and nested in a parallel \texttt{for}. Adding barriers immediately around \texttt{C} will result in parallel loop splitting immediately above and below \texttt{C}. As a result, the operations above and below \texttt{C} will be separated into their own parallel \texttt{for} and \texttt{C} will be the sole operation in the middle loop.
We can then apply one of the following techniques to interchange \texttt{C} with the parallel \texttt{for} thus making the latter immediate parent of the \texttt{barrier}.

\begin{figure}
    \centering

\begin{tabular}{@{}l|c@{}}
\begin{minipage}[T]{0.46\linewidth}
\begin{minted}[fontsize=\footnotesize, escapeinside=||]{mlir_lexer.py:MlirLexer -x}
parallel for %id=0 to N {
  for %j = 5 to 0 {
    if (%id < 2^%j)
      A[%id] += \
          A[%id + 2^%j]
    barrier
  }
}
\end{minted}
\end{minipage}&%
\begin{minipage}[T]{0.46\linewidth}
\begin{minted}[fontsize=\footnotesize, escapeinside=||]{mlir_lexer.py:MlirLexer -x}
for %j = 5 to 0 {
  parallel for %id=0 to N {
    if (%id < 2^%j)
      A[%id]+=A[%id + 2^%j]
    barrier
  }
}
\end{minted}
\end{minipage}
\end{tabular}
    \caption{\textbf{\emph{Left:}} A shared memory addition, which consists of a kernel call which contains for loop with a barrier inside. \textbf{\emph{Right:}} The same code with the barrier now directly within the parallel loop by performing an interchange of the parallel for loop with the serial for loop.}
    \label{fig:shmemreduction}
\end{figure}

Consider the case of a serial \texttt{for} loop containing a barrier, Fig.~\ref{fig:shmemreduction}.
This pattern is common in GPU code, e.g.,  to implement a reduction across threads~\cite{harris2007optimizing}.
As \texttt{barrier} must wait for all threads, each thread must execute the same number of \texttt{barrier}s.
Therefore, the number of iterations of the inner loop is the same for all threads, allowing for loop interchange. 

While an \texttt{if} statement can be considered a loop with zero or one iteration, directly interchanging it with the surrounding parallel \texttt{for} when necessary is more efficient.

\begin{figure}
    \centering

\begin{tabular}{@{}l|c@{}}
\begin{minipage}[T]{0.45\linewidth}
\begin{minted}[fontsize=\footnotesize, escapeinside=||]{mlir_lexer.py:MlirLexer -x}

parallel for %i=0 to N {
  do {
    run(%i)
    barrier
  } while(condition())
}

| |
\end{minted}
\end{minipage}&%
\begin{minipage}[T]{0.48\linewidth}
\begin{minted}[fontsize=\footnotesize, escapeinside=||]{mlir_lexer.py:MlirLexer -x}
%helper = alloca memref<i1>
scf.do {
  parallel for %i=0 to N {
    run(%i)
    barrier
    %c = condition()
    if %i == 0 {
      store %c, %helper[]
    }
  }
  %c = load %helper[]
} while(%c)
\end{minted}
\end{minipage}
\end{tabular}
    \caption{Parallel interchange around a \texttt{while} loop. As the \texttt{condition()} function call must be executed on each thread to preserve correctness, a helper variable is used which holds the value of the call on the first thread.}
    \label{fig:parallel_while}
\end{figure}

In MLIR, \texttt{for} loops must have their number of iterations computed prior to the loop. A \texttt{while} loop supports a dynamic exit condition. For example, consider the code in Fig.~\ref{fig:parallel_while}. To preserve correctness, we must execute the \texttt{condition()} call in every thread so a direct interchange would not be legal. However, the number of iterations must be equal across all threads due to the GPU synchronization semantics. Therefore, the interchange is possible thanks to a helper variable which stores the result of the condition from one thread that is used to decide whether another iteration is required. 

This illustrates one of the advantages of building such a system within MLIR/Polygeist. By being able preserve the high level structure of the program we can use more efficient patterns to remove barriers.

\changed{
\subsection{Usage}
CUDA transpilation with Polygeist is designed to be easy to use by allowing Polygeist to serve as a drop in replacement for an existing CUDA compiler like clang. Specifically, Polygeist extends the clang frontend, and as a result uses the same flags and syntax as clang. However, Polygeist also introduces several additional flags, such as \texttt{-cuda-lower} to specify the GPU-to-CPU translation and \texttt{-cpuify=XX} to specify a given method and set of parallel optimizations (see Section~\ref{sec:opt}) used for generating the CPU program.
}

\section{Parallel Optimization}
\label{sec:opt}

The high-level representation of both parallelism and GPU programs provided by Polygeist/MLIR enables a variety of optimizations. These include general optimizations that would apply to any parallel program as well as specific optimizations in the context of GPU to CPU conversion.

\subsection{Barrier Elimination \& Motion}
As GPU-style barriers have to be specially transformed to support CPU architectures, eliminating or simplifying any barriers can have dramatic effects. Moreover, even when running GPU code on the GPU, barrier elimination is a highly useful as any synchronization reduces parallelism. Much of the infrastructure for barrier elimination and simplification comes directly from its memory behavior defined in Section~\ref{sec:barrier_transform}. Given a barrier \texttt{B}, let $M^\dagger_{before}$ be the union of memory effects before \texttt{B} until either another barrier or the start of the parallel region, and let the union of memory effects after \texttt{B} until the end of the parallel region $M_{after}$. \changed{If there are no memory effects to the same location across the barrier other than a read-after-read (RAR) (i.e. $(M^\dagger_{before} \cap M_{after}) \setminus \mathrm{RAR} = \varnothing$)}, the barrier has its behavior subsumed by the prior barrier and can be eliminated. The symmetric condition $M_{before} \cap M^\dagger_{after} \setminus \mathrm{RAR} = \varnothing$ indicates that the barrier is subsumed by a subsequent barrier. A specific trivial case of eliminatable barrier is one that has no memory effects at all.

For example, consider the code in Fig.~\ref{fig:backprop}, which comes from the \texttt{backprop} Rodinia benchmark~\cite{rodinia}. The first and last \texttt{\_\_syncthreads} instructions are unnecessary. This can be proven from our memory-based barrier elimination algorithm above as follows. For the first barrier, $M_{before}$ (going all the way to the start) contains only a write to \texttt{node} and a read from \texttt{input}. $M^\dagger_{after}$ (going to the second \texttt{\_\_syncthreads}) contains a write to \texttt{weights} and a read from \texttt{hidden}. None of these conflict if, given the calling context, the pointers are known not to alias. 
Thus, it is safe to eliminate the barrier.

\begin{figure}
    \centering
\begin{minted}[fontsize=\footnotesize, escapeinside=||]{cuda}
__global__ void bpnn_layerforward(...) {
  __shared__ float node[HEIGHT];
  __shared__ float weights[HEIGHT][WIDTH];
  if ( tx == 0 )
   node[ty] = input[index_in] ;
  // Unnecessary Barrier #1
  __syncthreads();
  // Unnecessary Store #1
  weights[ty][tx] = hidden[index];
  __syncthreads();
   
  // Unnecessary Load #1
  weights[ty][tx] = weights[ty][tx] * node[ty];
  __syncthreads();   
   
  for ( int i = 1 ; i <= log2(HEIGHT) ; i++){
    if( ty % pow(2, i) == 0 )
      weights[ty][tx] += weights[ty+pow(2, i-1)][tx];
    __syncthreads();
  }

  hidden[index] = weights[ty][tx];
  // Unnecessary Barrier #2
  __syncthreads();

  if ( tx == 0 )
    output[by * hid + ty] = weights[tx][ty];
}
\end{minted}
    \caption{An example CUDA kernel from the Rodinia backprop test that contains unnecessary synchronization and unnecessary use of shared memory where a register would have sufficed.}
    \label{fig:backprop}
    \vspace*{-1em}
\end{figure}

The same memory analysis can also be applied to perform barrier motion. One simply needs to place a fictitious barrier at the intended location for a barrier to be moved to and check if the previous memory analysis would deduce that the current barrier is unnecessary, thereby permitting barrier motion.

\subsection{Memory-to-register promotion across barriers}
One of the goals of defining \texttt{barrier}'s semantics from its memory behavior is to enable memory optimizations to operate correctly and effectively in code that contains barriers. As described in Section~\ref{sec:barrier_transform}, barriers have the memory behavior of the code above and below them with the notable exception of the access from the current thread. This hole is important as it enables the memory-to-register promotion to operate on thread-local memory such as local variables. Moreover, this optimization is able to successfully replace slow memory reads with fast registers. For example, consider again the code in Figure \ref{fig:backprop}. Consider the load and store to \texttt{weights[ty][tx]} labeled ``Unnecessary Store \#1'' and ``Unnecessary Load \#1'', and the sync in between the two. The only value that can be loaded at that point is the same one which was stored earlier by the same read. Moreover, because that same location is overwritten before anyone else could read from \texttt{weights}, the first store can be safely eliminated once the load simply uses the register containing the value loaded from \texttt{hidden}. \changed{During the memory-to-register optimization, Polygeist can now successfully derive this forwarding property, since the hole in the memory properties described in Section~\ref{sec:barrier_transform} allows it to deduce that the barrier operation does not overwrite the store for the current thread.}
As a result traditional load and store forwarding correctly operate on the barrier code.

\subsection{Parallel loop-invariant code motion}
The traditional loop-invariant code motion optimization aims to move an instruction \texttt{I} outside serial "for" loops, reducing the number of times \texttt{I} is executed. If \texttt{I} may access memory, or has other side effects, in addition to checking that the operands of \texttt{I} are themselves loop invariant, the compiler must check that no other code within the "for" loop conflicts with the memory access performed by \texttt{I}.

On present compilers, while it is possible to apply loop-invariant code motion to serial for loops within GPU kernels, it is not possible to apply loop-invariant code motion to hoist instructions outside of a kernel call. This is in part due to the fact that GPU kernels are kept in a separate module from the CPU code which calls them, as well as a lack of understanding of parallelism in traditional compilers  (see Figure \ref{fig:intro_example}). 

Counter-intuitively, with the right semantics we can apply loop-invariant code motion to parallel for loops even if we would not be able to apply loop-invariant code motion to an equivalent serial loop. We will rely on the fact that semantics of our program permits us to arbitrarily interleave iterations of a parallel "for" loop as long as we maintain the orderings required by barriers. As such, it is legal to imagine running the program in lock-step. That is to say, if a parallel for loop had 10 instructions, each thread would execute instruction 1 before any thread would execute instruction 2, and so on. As a consequence, it is now legal to hoist an instruction so long as its operands are invariant and no \emph{prior} instruction in the parallel for loop conflicts with \texttt{I}. In other words, one does not need to check if \texttt{I} conflicts with any subsequent instruction in the parallel for loop to enable hoisting.

\begin{figure}
    \centering

\begin{tabular}{@{}l|c@{}}
\begin{minipage}[T]{0.48\linewidth}
\begin{minted}[fontsize=\footnotesize, escapeinside=||]{mlir_lexer.py:MlirLexer -x}
omp.parallel {
  omp.wsloop %i= 1 to 10 {
    codeA(%i)
  }
}
omp.parallel {
  omp.wsloop %i= 1 to 10 {
    codeA(%i)
  }
}
\end{minted}
\end{minipage}&%
\begin{minipage}[T]{0.48\linewidth}
\begin{minted}[fontsize=\footnotesize, escapeinside=||]{mlir_lexer.py:MlirLexer -x}
omp.parallel {
  omp.wsloop %i=1 to 10 {
    codeA(%i)
  }
  omp.barrier
  omp.wsloop %i=1 to 10 {
    codeA(%i)
  }
}
| |
\end{minted}
\end{minipage}
\end{tabular}
    \caption{Example of OpenMP parallel region fusion, as applied on MLIR. Given two adjacent OpenMP parallel regions, each of which initializes threads to be run with the given closure, fuse the two closures along with a thread barrier, thereby allowing the threads to be initialized once instead of twice.}
    \label{fig:omp_merge}
\end{figure}

\begin{figure}
    \centering

\begin{tabular}{@{}l|c@{}}
\begin{minipage}[T]{0.50\linewidth}
\begin{minted}[fontsize=\footnotesize, escapeinside=||]{cpp}
for (i=0; i<N; i++) {
  #pragma omp parallel for
  for (j=0; j<10; j++) {
    body(i, j);
  }
}

| |
\end{minted}
\end{minipage}&%
\begin{minipage}[T]{0.45\linewidth}
\begin{minted}[fontsize=\footnotesize, escapeinside=||]{cpp}
#pragma omp parallel
for (i=0; i<N; i++) {
  #pragma omp for
  for (j=0; j<10; j++) {
    body(i, j);
  }
  #pragma omp barrier
}
\end{minted}
\end{minipage}
\end{tabular}
    \caption{Example of OpenMP parallel region hoisting, as applied on C/C++. This is an extension of the OpenMP parallel region merging in Figure \ref{fig:omp_merge}, except as applied to an entire for loop. Rather than creating a separate closure / thread initialization for each iteration \texttt{i} of the outer loop, create the closure / thread initialization once. }
    \label{fig:omp_formerge}
\end{figure}

\subsection{Block Parallelism Optimizations}
\label{sec:block_opts}

OpenMP is our primary target for parallel execution on the CPU. It implements parallel "for" loops as two constructs. First, the loop is outlined into a function which is called once per thread, representing OpenMP's "parallel" construct. Then, within the outlined function, the iteration space is distributed across threads, representing OpenMP's "worksharing loop" construct. OpenMP also has a "barrier" construct, but with semantics \emph{different} than that of a GPU barrier.

When multiple parallel loops are executed in a row, e.g., following the barrier lowering from Section~\ref{sec:barrier_lowering}, the overhead of thread management can be reduced by fusing adjacent OpenMP "parallel" constructs~\cite{doerfert2018compiler} \emph{without} fusing the worksharing loops (see Fig.~\ref{fig:omp_merge}), thus not undoing the barrier lowering. Parallel region fusion can be extended a variety of constructs such as moving the OpenMP parallel region outside the surrounding "for" in Fig.~\ref{fig:omp_formerge}. This calls thread initialization once rather than $N$ times. Applying this generally to control flow constructs enables all of the parallel for loops generated by performing parallel loop fission on a block to have their OpenMP parallel (but not work sharing loops) fused.

As GPU programs tend to be written with high parallelism in mind, the parallelism provided by the different blocks may already saturate the number of available cores alone. If there is no use of shared memory, the block and thread parallelism can be collapsed into a single OpenMP parallel \texttt{for}, which will evenly divide the total iteration space in a single parallel region. However, if there is shared memory, our tool will generate nested parallel regions to represent the shared memory allocation. In this case, the additional overhead from the nested OpenMP parallel regions may outweigh the potential added parallelism. \changed{In addition, parallelizing the inner loops may lead to adverse memory effects such as false sharing, further penalizing performance~\cite{ppcg_spatial,vasilache2012joint}.} As such, we also support an optimization for serializing any nested OpenMP parallel regions. \changed{Performing such serialization may leverage memory locality to improve performance.}

\section{MocCUDA: Integration into PyTorch}
\label{sec:moccuda}

\subsection{Targeting CPU-only Supercomputers}
An aim of our work to model the GPU execution model in MLIR is the ability to execute high-performance codes originally written for GPUs on a supercomputer that has only CPUs, in particular, on the Fugaku machine with A64FX processors~\cite{fugaku}.
As a prime example, consider PyTorch~\cite{pytorch} that has not been successfully ported to the A64FX architecture.
PyTorch supports CPU execution using its "native" backend, which offers only a naive and thus poorly performing implementation for many kernels, e.g., a 6-way nested loop with no memory optimization for a 2D convolution.
On Intel CPUs, the execution of computationally-intensive kernels is delegated to the oneDNN library~\cite{oneDNN} which performs poorly on Arm CPUs as its tailored for common CPUs without high-bandwidth memory available on A64FX.
Fujitsu's fork of oneDNN~\cite{dnnl_aarch64} (which we also compare against in Section~\ref{ssec:moccudaeval}) required manual tuning to improve performance, and is still not universally competitive with GPUs.
Interestingly, CPUs with high-bandwidth memory can benefit from computation organization similar to that of GPUs.
We therefore design ``MocCUDA'' to mock the GPU backed for Pytorch by transpiling CUDA kernels into OpenMP-enabled parallel CPU kernels using the mechanism described in the previous sections.

\subsection{Architecture of MocCUDA and Integration of Polygeist}\label{ssec:moccudaarch}

We implemented a compatibility layer, MocCUDA, to enable transparent execution of PyTorch GPU backend exclusively on CPU with the goal of avoiding any manual re-engineering and interoperability with existing libraries. While PyTorch does have its own CPU backend, its design is poorly compatible with massively parallel computers running CPUs. In particular, performance issues are caused by synchronous kernel execution, mismatched memory layouts, and a presumption of low-speed memory.

Rather than attempting a challenging and time-consuming redesign the PyTorch CPU backend, we decided to base MocCUDA on the design of PyTorch's GPU backend given the relative similarity of the platform to A64FX. Therefore, MocCUDA must handle and replace four types of operations:
\begin{enumerate}
    \item calls to the CUDA runtime (CUDART),
    \item calls to deep-learning specific functions in cuDNN,
    \item calls to auxiliary CUDA libraries, and
    \item executions of CUDA kernels from within PyTorch.
\end{enumerate}
Only the latter can be compiled since the rest of the functions are distributed as binary libraries.

To minimize the scope of work for the prototype, we profiled all interactions of PyTorch with the CUDA backend to identify (1)--(4) for a well-known deep
learning problem,
namely the ResNet-50~\cite{he2016deep} neural network.

From profiling, PyTorch's interaction with the CUDART is mostly limited to identifying properties of installed GPUs, memory management, and management and
synchronization of CUDA streams. For the prototype, we limit ourselves to emulating one GPU per
NUMA node, and only managing explicit memory allocation and data transfer.
To give PyTorch access to GPU device properties,
we dump the data of a real GPU, Nvidia GeForce RTX 2080 Ti, into a file which is used
later by MocCUDA on the system without GPU. We emulate CUDA streams through Apple's Grand Central
Dispatch (GCD)~\cite{sakamoto2012grand} which enables asynchronicity of the emulated GPU operation from Pytorch's management layers.

The ResNet-50 model only exercises a subset of closed-source cuDNN functions (forward and backward passes for the convolutional layers, batch normalization layers, and tensor additions). We re-implemented all necessary cuDNN
function variations with focus on OpenMP-parallelization and HBM-friendly Im2Col plus GEMM convolutions
as well as support for the NCHW layout \changed{(batch N, channels C, height H, width W)}. Furthermore, we identified calls to the cuBLAS library, and
implemented wrapper functions to intercept these calls and dispatch them instead to the BLAS library
designed for CPUs, such as MKL, OpenBLAS, or Fujitsu's SSL2~\cite{fujitsu_SSLII} for A64FX. A similar approach
is possible for other libraries linked into PyTorch, such as cuFFT, but is not necessary for ResNet-50.

Besides CUDART and other library calls, we observed \changed{\emph{custom kernels}, i.e. kernels implemented in CUDA within PyTorch rather than provided by (binary) libraries,} and identified the high-level functions which invoke them.
Functions required by ResNet-50 include strided tensor kernels (add, multiply, etc), aggregation operations like ``Softmax'', among others. The negative log-likelihood loss kernel uses CUDA's \texttt{\_\_syncthreads()}
barriers. Porting these CUDA-based PyTorch functions to CPU involves labor-intensive re-engineering
whereas Polygeist performs automatic translation. For demonstration purposes, we use
Polygeist to automatically translate the \changed{\texttt{ClassNLLCriterion\_updateOutput}, which uses \texttt{\_\_syncthreads()}, and \texttt{ClassNLLCriterion\ \_updateGradInput}} functions\changed{, and the functions transitively called from these,} from CUDA to OpenMP, and integrate the result into MocCUDA.

We combine all the above described wrappers and re-implementations into MocCUDA,
which can be used with \texttt{LD\_PRELOAD} to intercept CUDART/cuDNN calls to train ResNet-50 with the CUDA backend on CPUs.

\section{Evaluation}\label{sec:evaluation}

We demonstrate the advantages and applicability of our approach on two well-known GPU benchmark suites: a subset of the GPU Rodinia benchmark suite~\cite{rodinia} and a PyTorch implementation of a Resnet-50 neural network. These benchmarks were chosen to 1) provide a rough performance comparison of our GPU to CPU compilation on a benchmark suite (Rodinia) that has hand-coded CPU versions and 2) demonstrate a successful end-to-end integration of our system into a useful and real application (PyTorch Resnet-50) on Supercomputer Fugaku, which does not have any GPUs. Additionally, we compare the performance of our approach to the existing MCUDA~\cite{mcuda} tool on a CUDA matrix multiplication.

\begin{figure}
    \centering
    \includegraphics[width=\columnwidth]{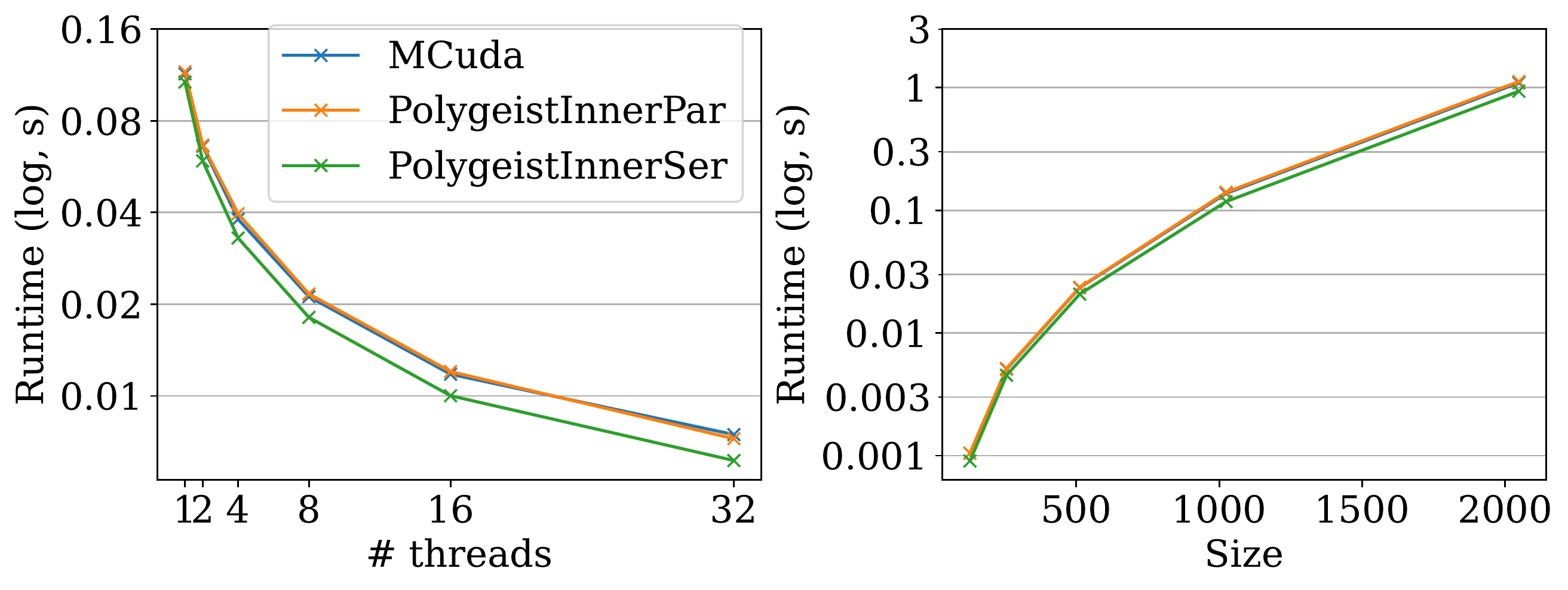}
    \vspace*{-2em}
    \caption{\change{PolygeistInnerPar has a similar performance to MCUDA, while PolygeistInnerSer outperforms MCUDA. PolygeistInnerSer disables inner loop parallelization similaly to MCUDA, whereas PolygeistInnerPar keeps both the blocks and threads parallel. Left: Average runtime as a function of thread count (averaging over matrix sizes). Right: Average runtime as a function of matrix size (averaging over thread counts).}}
    \label{fig:mcuda}
    \vspace*{-1em}
\end{figure}

For Rodinia, we compare our translated CUDA to CPU code against OpenMP versions of the benchmarks, where they exist, as well as a run on a GPU. For the PyTorch Resnet-50, we compare against the ``native'' and oneDNN backends.

Polygeist\footnote{\changed{Relevant versions of MocCUDA and Polygeist are available at \url{https://anonymous-data.s3.amazonaws.com/MocCUDA-master.zip} and \url{https://anonymous-data.s3.amazonaws.com/Polygeist-main.zip}.
}} was compiled against LLVM 15 at commit \texttt{89525cbf}.
For the PyTorch Resnet-50, we compile Pytorch v1.4.0 using Nvidia's CUDA 11.6 SDK for Arm\footnote{Even though we will run PyTorch on a GPU-less system, we must compile PyTorch on a CUDA-enabled system to ensure the correct code is emitted. We also prevented inlining of three Pytorch functions.}, LLVM 13, and Fujitsu's SSL2 v1.2.34 library. For the baseline PyTorch measuremets% 
, we use Fujitsu's pre-installed PyTorch (v1.5.0).

We evaluate the Rodinia and matrix multiplication tests on an AWS c6i.metal instance (dual-socket Intel Xeon Platinum 8375C CPU at 2.9~GHz with 32 cores each and 256~GB RAM) running Ubuntu 20.04. Measurements were performed on the first socket, with hyperthreading and turbo boost disabled. Each number is the median of at least 5 repetitions.

\subsection{Comparison to MCUDA}\label{sec:mcuda}
First, we compare our approach to the previous work in MCUDA~\cite{mcuda}%\todo{Precise version description}
. MCUDA is an AST-level tool which produces new CPU C/C++ as an output and uses a similar loop fission technique to handle synchronization. As a source-to-source tool, it handles only a fraction of the input language, making it unable to handle Rodinia programs.
Instead, we compare the runtimes of a matrix multiplication kernel across a range of threads (1--24) and matrix sizes ($128\!\times\!128$ -- $2048\!\times\!2048$) in Fig.~\ref{fig:mcuda}.
\change{Polygeist with all optimization excluding serialization of the inner loop (PolygeistInnerPar) produces code within $1.3\%$ of MCUDA on average. Specifically PolygeistInnerPar has a $1.5\%$ slowdown on 1 thread, and $3.2\%$ speedup on 32 threads. This behavior is caused by OpenMP overhead in handling nested parallel constructs. In fact, MCUDA only parallelizes the outermost loop. When
Polygeist serializes the inner loops (PolygeistInnerSer), it achieves an overall $14.9\%$ speedup over MCUDA, with a $4.5\%$ speedup on 1 thread and $21.7\%$ speedup on 32 threads.}

\subsection{Use case 1: Rodinia Benchmarks}\label{ssec:rodinia}

\begin{figure*}
    \centering
    \includegraphics[width=\linewidth]{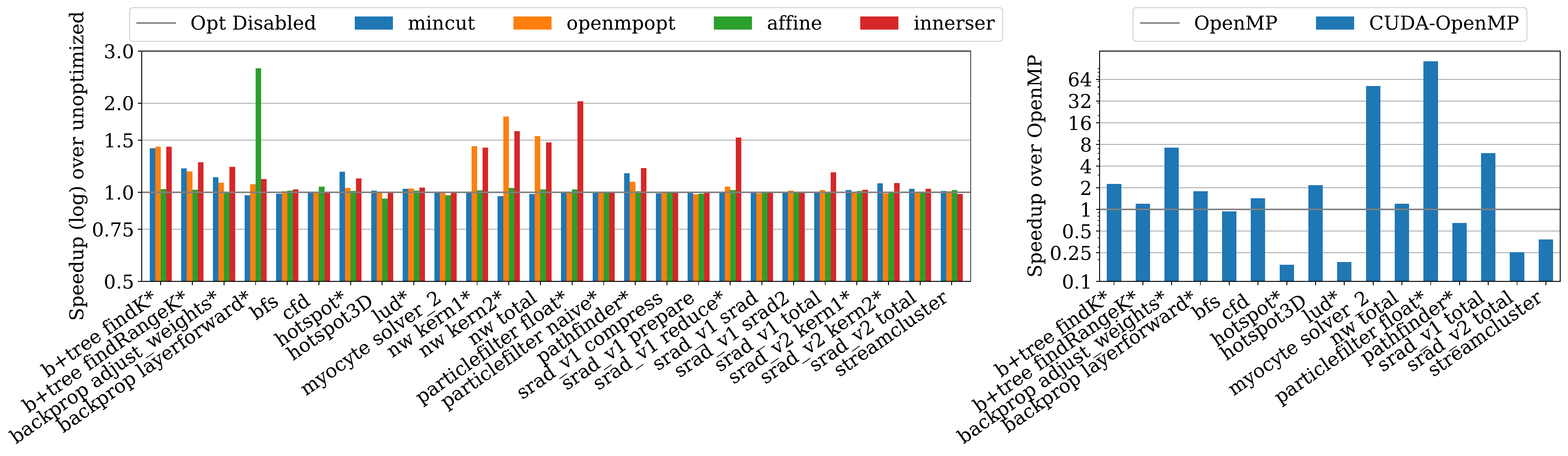}
    \vspace*{-2em}
    \caption{Left: Relative speedup (higher is better) of kernels with various parallel and/or CPU optimizations applied. Right: Speedup of Rodinia CUDA code when compiled to OpenMP compared against native Rodinia OpenMP code (when available). Benchmarks containing barriers are marked with an asterisk.}
    \label{fig:rodinia_speedups}
\end{figure*}

We benchmarked a subset of 14 benchmarks that are currently supported by Polygeist, and had a nontrivial runtime.\footnote{\changed{The \texttt{hybridsort}, \texttt{kmeans}, \texttt{leukocyte}, \texttt{mummergpu} \texttt{huffman} and \texttt{heartwall} use unsupported C++ or CUDA features within Polygeist (virtual functions and texture memory). The \texttt{lavaMD} and \texttt{dwt2d} benchmarks use \emph{ill-formed} C++ with undefined behavior due to reading from uninitialized memory (GPU driver zero-initializes shared memory, but is not required to do so). } The \texttt{nn} and \texttt{gaussian} tests ran in $\leq0.005$ seconds.}
We verified correctness of our flow by comparing the program outputs produced by compiling with \texttt{nvcc} and executed on a GPU, and compiled by our flow and executed on a CPU. We inserted timing measurements across kernels and/or computational portions of the code that include kernels, in some cases multiple per benchmark. Where possible, we time equivalent portions of the OpenMP versions of the same benchmarks.\footnote{For \texttt{nn}, already excluded due to its trivial runtime, the two versions differ in data loading (dynamically as it is run, in the OpenMP code, and preloading all data, in the CUDA code) and ought to be excluded for this reason as well.}

We compare the Rodinia CUDA benchmarks compiled for the CPU with the Rodinia OpenMP verions of the benchmark in Fig.~\ref{fig:rodinia_speedups}(right). While there is some variation from benchmark to benchmark, overall our approach is on par with the hand-coded versions of the benchmarks, and even nets a 76\% geomean performance improvement, when the inner serialization optimization is enabled. Without inner serialization, we still see a geomean speedup of 43.7\%. Some benchmarks such as \texttt{hotspot} and \texttt{pathfinder} employ optimizations techniques for stencil computations which duplicate computation across threads in order to reduce synchronization overhead and make better use of the parallelism available in a GPU. This makes the CUDA code significantly more complex than the OpenMP version which causes them to  perform worse. \changed{The CUDA-OpenMP versions of \texttt{lud} and \texttt{srad\_v2} tests are slower as the program performs additional work to cache data within shared memory. The CUDA-OpenMP version of \texttt{particlefilter} receives a relative speedup as the pure OpenMP version of the code achieves the desired dependency structure through separate OpenMP ``parallel for'' loops, whereas in the CUDA code, this was done with a \texttt{\_\_syncthreads}. Polygeist was able to successfully optimize code surrounding the barrier, resulting in the speedup. The speedup for \texttt{backprop} is partially due to parallel optimizations (see Fig.~\ref{fig:rodinia_speedups}(left)) and partially due to the CUDA code being implemented with a linear array, as required by CUDA, instead of the double-pointer used in the OpenMP code. The \texttt{myocyte} and \texttt{sradv1}, both achieve speedups due to code optimization across the parallel region boundary, as well as inner serialization.  }

\begin{figure}
    \centering
    \includegraphics[width=.95\columnwidth]{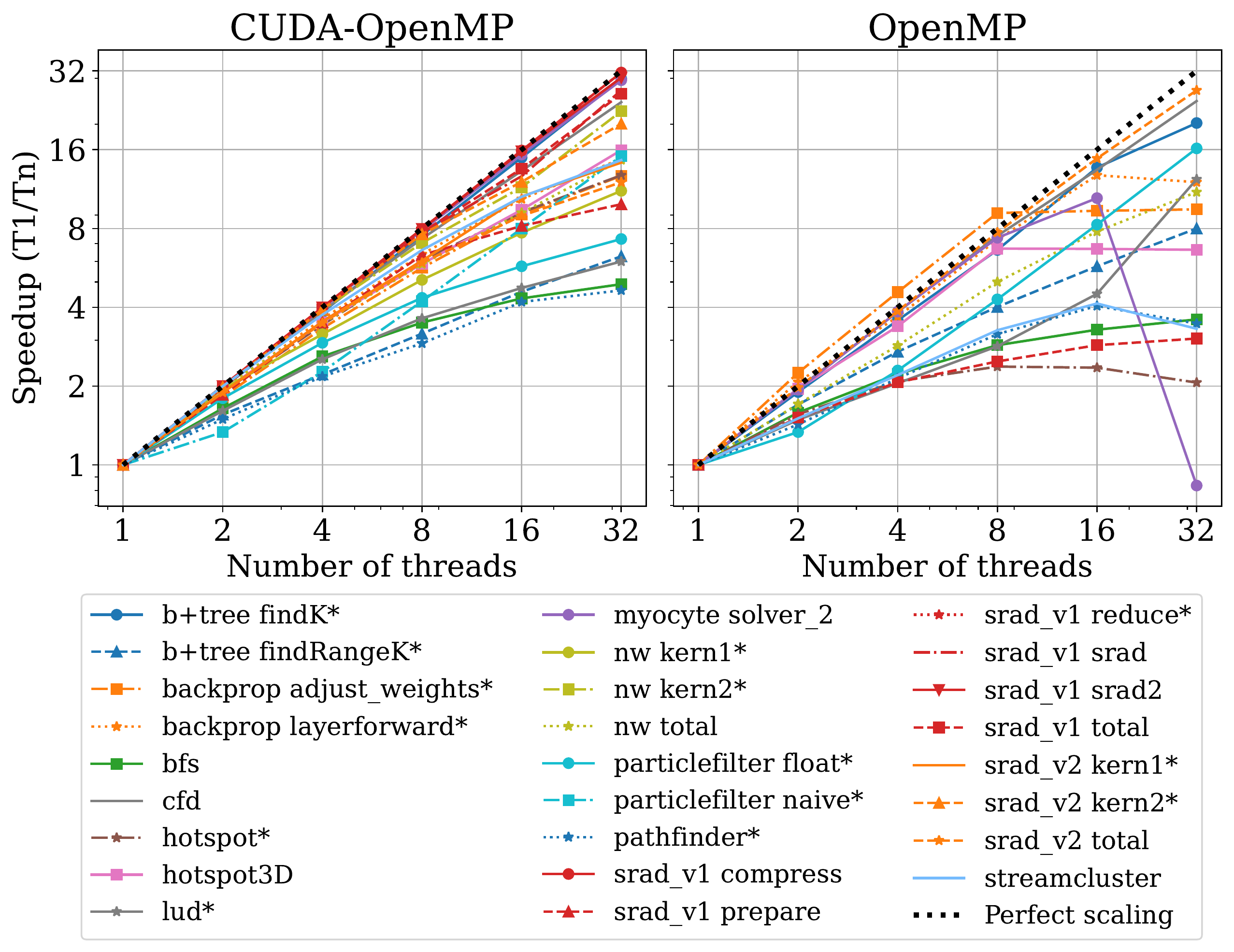}
    \vspace*{-1em}
    \caption{Scaling behavior behavior of CUDA Rodinia kernels, when run on the CPU with OpenMP, and OpenMP Rodinia kernels (where available), using 32 threads. Not all Rodinia CUDA kernels have OpenMP versions.}
    \label{fig:polygeist_scaling}
    \vspace*{-1em}
\end{figure}

We test the scaling properties of our approach by comparing transpiled CUDA with native OpenMP kernels in Fig.~\ref{fig:polygeist_scaling}. Transpiled CUDA codes generally scale much better than the native OpenMP versions. As most CUDA programs are written with thousands of threads in mind, this indicates that our framework was able to preserve that parallelism as the GPU-specific constructs were being rewritten for CPU-compatible equivalents. On 32 threads without inner serialization, transpiled CUDA codes had a geomean speedup of $16.1\times$ across all tests. As OpenMP versions of benchmarks do not exist for all tests, if we consider only CUDA codes for which there exists an OpenMP version, we find a geomean speedup of $14.0\times$, whereas OpenMP has only a speedup of $7.1\times$. Serializing the inner loop slightly reduces scalability, but still results in improved scalability over OpenMP, finding a geomean speedup of $14.9\times$ over all tests with inner serialization enabled, and a $12.5\times$ speedup on codes with OpenMP versions.

We perform an ablation analysis to study how individual optimizations impact performance. The ``mincut'' series in Fig.~\ref{fig:rodinia_speedups}(left) shows performance measurements for our approach with the optimization outlined in Section~\ref{sec:loopsplitting} to reduce the amount of data preserved across barriers. This is only relevant for benchmarks containing barriers (marked by an asterisk in the Figure). On applicable benchmarks, mincut provides a 4.1\% geomean speedup.
The ``openmpopt'' series in Fig.~\ref{fig:rodinia_speedups}(left) demonstrates the impact of OpenMP region merging and similar optimizations and results in a 8.9\% geomean speedup.
The ``affine'' series in Fig.~\ref{fig:rodinia_speedups}(left) shows the result of raising control flow to their affine variants and enabling simple loop optimizations (such as loop unrolling). While this produces a geomean speedup of 4.6\% across the board, it results in a $2.6\times$ speedup for the backprop layerforward test as it results in a loop containing synchronization being fully unrolled and reduced to \texttt{if} statements.

\subsection{Use case 2: Pytorch/Resnet50 Test}
\label{ssec:moccudaeval}
To evaluate the PyTorch Resnet-50, we execute 
a full node-parallel training run on one TofuD unit of the Fugaku FX1000 supercomputer, comparing against the native PyTorch CPU backend and the optimized oneDNN backend, as available.

We ran multiple forward and back propagation passes of Resnet-50 on 224$\times$224 ImageNet in a data-parallel fashion.
We employ Horovod's synthetic benchmarking script (configured for the Resnet-50 neural network model)~\cite{horovod}. We build Horovod v0.19.5 with CUDA SDK, LLVM, and Fujitsu's MPI library to enable multi-node, distributed deep learning on top of Pytorch. 
We assign one MPI rank per A64FX core memory group (CMG), emulating up to 4 GPUs per node, and scale the test from one node (2 ranks) to 12 nodes (48 ranks) in one TofuD unit (smallest 2$\times$3$\times$2 torus) while keeping the number of OpenMP threads fixed at 12 to accommodate one thread per core. We use Pytorch v1.4.0 for our approach, while the other backends depend on Pytorch v1.5.0.
Performance measurements were taken using Benchmarker~\cite{benchmarker}, which
orchestrates sets up the neural network via torchvision, creates images, executes the layer using PyTorch, and returns an images/s throughput metric. 
We run with batch sizes \changed{1--12 on 1--64 threads}, averaging across epochs.

\begin{figure}[tbp]
    \includegraphics[width=\columnwidth]{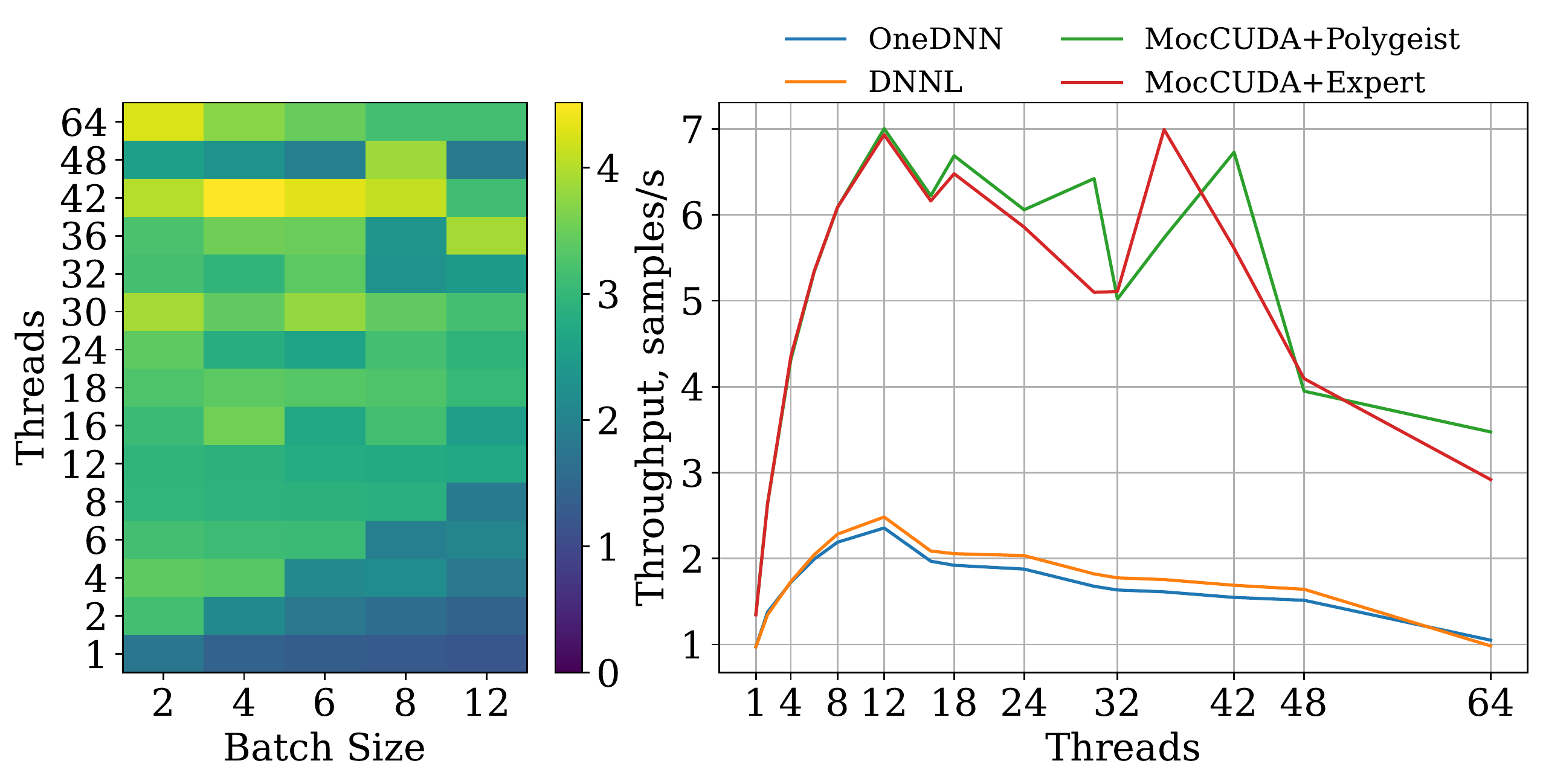}
    \vspace*{-2em}
    \caption{\changed{ResNet50 training on Fugaku node. Left: heatmap of relative throughput increase of ``MocCUDA+Polygeist'' over Fujitsu-\emph{tuned} oneDNN, higher is better. Right: geomean throughput across batch sizes; ``MocCUDA+Expert'' uses an expert-written OpenMP kernel; ``MocCUDA+Polygeist'' uses the generated kernel.}}
    \label{fig:resnet50}
    \vspace*{-1em}
\end{figure}

\changed{We observed that MocCUDA systematically outperforms Fujitsu's tuned oneDNN across batch sizes and thread counts, yielding up to 4.5$\times$ throughput increase (geomean 2.7$\times$, min 1.2$\times$) as shown in Fig.~\ref{fig:resnet50}. MocCUDA with our expert-written kernels is comparable to MocCUDA with the kernels generated by Polygeist, described in Section~\ref{ssec:moccudaarch}.}

The improvement can be explained by a combination of the PyTorch CPU design and performance characteristics of oneDNN described in Section~\ref{ssec:moccudaarch}. As Intel's oneDNN~\cite{oneDNN} does not account for HBM available on A64FX, it uses cache-friendly direct convolutions instead of GEMM-based convolutions, less efficient in presence of HBM for Arm CPUs. While the custom fork of oneDNN tuned by Fujitsu~\cite{dnnl_aarch64}, improves upon Intel oneDNN's performance (though by a geomean of $6\%$), it still leaves room for performance improvements.

This demonstrates that our approach is capable of automatically deriving efficient versions of deep learning kernels (and potentially other applications) from their CUDA versions, thus addressing the limitations of missing or inefficient kernels for CPUs with high-bandwidth memory without the need for reverse or re-engineering the application.

\section{Related Work}
\label{sec:related_work}
\subsection{GPU to CPU Synchronization}
One of the first tools for emulating GPUs on a CPU was provided directly by NVIDIA for debugging purposes and emulated each thread on the GPU with a distinct CPU thread.
While functional, the large gap in the number of available threads makes the emulation inefficient.

MCUDA~\cite{mcuda} (2008) performs an AST transformation of C GPU code to generate new C CPU code that calls a thread-independent parallel \texttt{for} routine. MCUDA pioneered the use of ``deep fission'' to handle synchronization, which splits parallel loops and other constructs at synchronization points in order to eliminate them. This fission technique is also applied in other tools: Ocelot~\cite{diamos2010ocelot} (2010), a binary-translation tool that parses PTX assembly into LLVM and just-in-time compiles kernel functions; POCL~\cite{jaaskelainen2015pocl} (2015), a Clang/LLVM compiler pass for OpenCL; COX~\cite{han2021cox} (2021), another LLVM transformation pass for translation of CUDA that uses fission, and notably handles warp-level primitives; and even this work. \changed{While the intuition behind the fission approach is similar to that used here, we apply fission inside of a high-level compiler, rather than either source or a low-level IR. As demonstrated in Section~\ref{sec:barrier_transform}, performing fission on structured programs enables more efficient code transformations. While performing fission at a source-level misses the opportunity to run optimizations (such as barrier elimination) before fission and applying fission at a low-level requires attempting to reconstruct the high-level structure, operating within MLIR allows us to both apply optimization and preserve high-level structure. Moreover, source-level tools tend to be quite fragile as they must re-implement parsing and semantics or the target language (e.g. C++), and as a result only operate on a limited subset of the input language,  requiring re-engineering effort to replace unsupported constructs (like pointer arithmetic).
}

Another approach uses continuation-passing to handle synchronization by creating state machine of all synchronization points (e.g. ``microthreading'')~\cite{stratton2010efficient} (2010).  Karrenberg and Hack~\cite{karrenberg2012improving} (2012) propose a continuation-passing approach in LLVM that includes an algorithm for detecting and reducing divergence in the control-flow-graph, with follow up work to minimizing live values to reduce memory traffic~\cite{DBLP:conf/cc/MollDH16}.

VGPU~\cite{DBLP:conf/icppw/PatelTDC21} (2021) is Similar to NVidia's original virtual GPU, except now uses C++ \texttt{std::thread} and performs synchronization using \texttt{std::atomic\_thread\_fence}. Shared memory, implemented as a single global in LLVM, is expanded by the number of blocks.

\looseness=-1
Prior work that operates at the low-level LLVM IR extends significant effort to reconstruct high-level constructs, such as loops and kernel configurations, required for either efficient fission or continuation passing. For example, POCL~\cite{jaaskelainen2015pocl} runs various canonicalizations and loop transformations to rewrite the control flow graph and attempt to recognize it as one of several forms that can be handled.
Prior work that operates at source/AST level (e.g. MCUDA), beyond still needing to recognize GPU-level concepts, 
cannot benefit from optimizations that simplify the code resulting in easier control flow. 

In contrast, by operating on MLIR's mix-of-abstractions, we are able to simultaneously preserve source-level structure and perform program transformations such as loop unrolling or LICM motion that can, e.g., remove nested synchronization.

\subsection{Parallel Portablity/IR, \& OpenMP Optimizations}
Several tools define new abstractions in the host language that are naturally amenable to CPU or GPU execution. Examples include ISPC~\cite{pharr2012ispc}, \changed{RAJA~\cite{beckingsale2019performance}, Kokkos~\cite{edwards2014kokkos}, or MapCG~\cite{hong2010mapcg} (limited to map-reduce computation)} in C++, Loo.py~\cite{10.1145/2627373.2627387} in Python, and KernelAbstractions.jl~\cite{valentin_churavy_2022_6324344} in Julia. These approaches provide performance portability for any new code written with them. However, any existing code must be rewritten in said framework (and may or may not compose with code written in other frameworks or other languages).

\change{
Several pieces of prior art discuss intermediate representations for parallelism, such as Tapir~\cite{schardl2019tapir} for representing Cilk~\cite{frigo1998implementation} in LLVM; OpenMPIR~\cite{openmpir} for representing OpenMP in LLVM, PPIR~\cite{schmitz2021ppir} for pattern trees, and the MLIR OpenMP Dialect; as well as SDf3~\cite{stuijk2006sdf} for visually representing concurrency as a control-flow graph. These works primarily focus on the \emph{representation} for their particular style of parallelism (e.g. OpenMP tasks in OpenMPIR), which does not include GPU-style barriers,  rather than on parallel-specific \emph{transformations} (such as barrier elimination) or optimizations, with the exception of consistency/race checks or automatic parallelization~
\cite{moon1999evaluation, oancea2012logical}.}

The use of OpenMP parallel region expansion is known to be beneficial~\cite{doerfert2018compiler}.
Clang/LLVM optionally supports the transformation in a weaker form, namely merging of OpenMP parallel regions in the same control level~\cite{openmpopt}.

\subsection{Barriers}
Several pieces of prior work have explored the semantics of barrier or synchronization instructions, including as it relates to GPUs. Work has been done to verify the correctness of barriers~\cite{10.1145/268946.268974}. \cite{sorensen2021specifying} explores and experimentally evaluates the forward progress / fairness models of various GPU vendors. \cite{sorensen2016portable} implements a GPU barrier operation that applies across work-groups, as opposed to just within a work group. \changed{\cite{sura2005compiler} add Java memory barriers to programs to ensure weak and sequential consistency semantics. They find that without synchronization and delay set analysis, introducing consistency semantics has an average $26.5\times$ slowdown, whereas when using these analyses to insert fewer syncrhonzations can achieve a 10\% and  26\% slowdown for weak and sequential consistency, respectively.}

\section{Conclusion}
\label{sec:conclusion}

By extending Polygeist/MLIR, we have developed an end-to-end system capable of representing, optimizing, and transpiling CPU and GPU parallel programs. Being able to simultaneously represent and convert between distinct parallel frameworks is crucial as HPC increasingly relies on (heterogeneous) parallelism for its workloads. A key component of our framework is the development of a high-level barrier operation, key to representing GPU programs, whose semantics can be fully defined by its memory behavior. Unlike prior representations of parallel barriers, our semantics enable direct integration of barriers within optimization. To validate the efficacy of our approach we demonstrate GPU to CPU optimization and transpilation of a subset of the Rodinia benchmark suite on a commodity CPU and transcompile an efficient Resnet-50 from the PyTorch CUDA source to be run on the CPU-only Supercomputer Fugaku. \changed{While there is case-by-case variance due to differences in implementation between CPU and GPU, the Rodinia benchmark suite achieves a 76\% geomean speedup of the transpiled GPU code over handwritten OpenMP versions. Similarly, a $\approx2\times$ speedup of CUDA PyTorch kernels above the native PyTorch CPU backend can be observed }

\changed{Currently, the transpiled GPU code keeps the same schedule when run on the CPU, except for the innermost loop serialization that improves performance. A fruitful avenue of future work may perform advanced rescheduling the code to better take advantage of CPU-style memory hierarchies.}

\iftoggle{releaseStuffAfterDoubleBlind}{%

\section*{Acknowledgment}
\label{sec:ackno}

Thanks to Valentin Churavy of MIT and Albert Cohen of Google for thoughtful discussions about transformations within MLIR. Thanks to Douglas Kogut, Jiahao Li, and Bojan Serafimov for thoughtful discussions about parallel optimizations within the compiler.

William S. Moses was supported in part by a DOE Computational Sciences Graduate Fellowship DE-SC0019323, in part by Los Alamos National Laboratories grant 531711, and in part by the United States Air Force Research Laboratory and the United States Air Force Artificial Intelligence Accelerator and was accomplished under Cooperative Agreement Number FA8750-19-2-1000. Johannes Doerfert was supported in part by the Applied Mathematics activity within the U.S. Department of Energy, Office of Science, Advanced Scientific Computing Research Program, under contract number DE-AC02-06CH11357; in part by the Exascale Computing Project (17-SC-20-SC), a collaborative effort of two U.S. Department of Energy organizations (Office of Science and the National Nuclear Security Administration) responsible for the planning and preparation of a capable exascale ecosystem, including software,applications, hardware, advanced system engineering, and early testbed platforms, in support of the nation's exascale computing imperative.
This work was supported in part by the Japan Society for the Promotion of Science KAKENHI Grant
Number 19H04119 and by the Japanese New Energy and Industrial Technology Development
Organization (NEDO).

The views and conclusions contained in this document are those of the authors and should not be interpreted as representing the official policies, either expressed or implied, of the United States Air Force or the U.S. Government. The U.S. Government is authorized to reproduce and distribute reprints for Government purposes notwithstanding any copyright notation herein.

}

%\bibliographystyle{IEEEtran}
%\bibliography{bibliography}

\begin{thebibliography}{10}
\providecommand{\url}[1]{#1}
\csname url@samestyle\endcsname
\providecommand{\newblock}{\relax}
\providecommand{\bibinfo}[2]{#2}
\providecommand{\BIBentrySTDinterwordspacing}{\spaceskip=0pt\relax}
\providecommand{\BIBentryALTinterwordstretchfactor}{4}
\providecommand{\BIBentryALTinterwordspacing}{\spaceskip=\fontdimen2\font plus
\BIBentryALTinterwordstretchfactor\fontdimen3\font minus
  \fontdimen4\font\relax}
\providecommand{\BIBforeignlanguage}[2]{{%
\expandafter\ifx\csname l@#1\endcsname\relax
\typeout{** WARNING: IEEEtran.bst: No hyphenation pattern has been}%
\typeout{** loaded for the language `#1'. Using the pattern for}%
\typeout{** the default language instead.}%
\else
\language=\csname l@#1\endcsname
\fi
#2}}
\providecommand{\BIBdecl}{\relax}
\BIBdecl

\bibitem{fugaku}
M.~Sato, Y.~Ishikawa, H.~Tomita, Y.~Kodama, T.~Odajima, M.~Tsuji, H.~Yashiro,
  M.~Aoki, N.~Shida, I.~Miyoshi, K.~Hirai, A.~Furuya, A.~Asato, K.~Morita, and
  T.~Shimizu, ``Co-design for {{A64FX}} manycore processor and
  {{``Fugaku''}},'' in \emph{SC20: International Conference for High
  Performance Computing, Networking, Storage and Analysis}, 2020, pp. 1--15.

\bibitem{pytorch}
A.~Paszke, S.~Gross, F.~Massa, A.~Lerer, J.~Bradbury, G.~Chanan, T.~Killeen,
  Z.~Lin, N.~Gimelshein, L.~Antiga \emph{et~al.}, ``Pytorch: An imperative
  style, high-performance deep learning library,'' \emph{Advances in neural
  information processing systems}, vol.~32, 2019.

\bibitem{oneDNN}
\BIBentryALTinterwordspacing
Intel, ``Oneapi-src/onednn: Oneapi deep neural network library (onednn).''
  [Online]. Available: \url{https://github.com/oneapi-src/oneDNN}
\BIBentrySTDinterwordspacing

\bibitem{dnnl_aarch64}
\BIBentryALTinterwordspacing
Fujitsu. [Online]. Available: \url{https://github.com/fujitsu/dnnl_aarch64}
\BIBentrySTDinterwordspacing

\bibitem{opencl}
\BIBentryALTinterwordspacing
P.~Du, R.~Weber, P.~Luszczek, S.~Tomov, G.~Peterson, and J.~Dongarra, ``From
  {{CUDA}} to {{OpenCL}}: Towards a performance-portable solution for
  multi-platform gpu programming,'' \emph{Parallel Computing}, vol.~38, no.~8,
  pp. 391--407, 2012. [Online]. Available:
  \url{https://www.sciencedirect.com/science/article/pii/S0167819111001335}
\BIBentrySTDinterwordspacing

\bibitem{openacc}
J.~A. Herdman, W.~P. Gaudin, O.~Perks, D.~A. Beckingsale, A.~C. Mallinson, and
  S.~A. Jarvis, ``Achieving portability and performance through {{OpenACC}},''
  in \emph{2014 First Workshop on Accelerator Programming using Directives},
  2014, pp. 19--26.

\bibitem{kokkos}
\BIBentryALTinterwordspacing
H.~{Carter Edwards}, C.~R. Trott, and D.~Sunderland, ``Kokkos: Enabling
  manycore performance portability through polymorphic memory access
  patterns,'' \emph{Journal of Parallel and Distributed Computing}, vol.~74,
  no.~12, pp. 3202--3216, 2014, domain-Specific Languages and High-Level
  Frameworks for High-Performance Computing. [Online]. Available:
  \url{https://www.sciencedirect.com/science/article/pii/S0743731514001257}
\BIBentrySTDinterwordspacing

\bibitem{spiral}
F.~Franchetti, T.~M. Low, D.~T. Popovici, R.~M. Veras, D.~G. Spampinato, J.~R.
  Johnson, M.~Püschel, J.~C. Hoe, and J.~M.~F. Moura, ``Spiral: Extreme
  performance portability,'' \emph{Proceedings of the IEEE}, vol. 106, no.~11,
  pp. 1935--1968, 2018.

\bibitem{halide}
\BIBentryALTinterwordspacing
J.~Ragan-Kelley, C.~Barnes, A.~Adams, S.~Paris, F.~Durand, and S.~Amarasinghe,
  ``Halide: A language and compiler for optimizing parallelism, locality, and
  recomputation in image processing pipelines,'' in \emph{Proceedings of the
  34th ACM SIGPLAN Conference on Programming Language Design and
  Implementation}, ser. PLDI '13.\hskip 1em plus 0.5em minus 0.4em\relax New
  York, NY, USA: Association for Computing Machinery, 2013, p. 519–530.
  [Online]. Available: \url{https://doi.org/10.1145/2491956.2462176}
\BIBentrySTDinterwordspacing

\bibitem{tc}
\BIBentryALTinterwordspacing
N.~Vasilache, O.~Zinenko, T.~Theodoridis, P.~Goyal, Z.~Devito, W.~S. Moses,
  S.~Verdoolaege, A.~Adams, and A.~Cohen, ``The next 700 accelerated layers:
  From mathematical expressions of network computation graphs to accelerated
  gpu kernels, automatically,'' \emph{ACM Trans. Archit. Code Optim.}, vol.~16,
  no.~4, oct 2019. [Online]. Available: \url{https://doi.org/10.1145/3355606}
\BIBentrySTDinterwordspacing

\bibitem{mcuda}
\BIBentryALTinterwordspacing
J.~A. Stratton, S.~S. Stone, and W.-m.~W. Hwu, ``{MCUDA}: An efficient
  implementation of {CUDA} kernels for multi-core {CPUs},'' in \emph{Languages
  and Compilers for Parallel Computing}, J.~N. Amaral, Ed.\hskip 1em plus 0.5em
  minus 0.4em\relax Springer Berlin Heidelberg, 2008, vol. 5335, pp. 16--30,
  series Title: Lecture Notes in Computer Science. [Online]. Available:
  \url{http://link.springer.com/10.1007/978-3-540-89740-8_2}
\BIBentrySTDinterwordspacing

\bibitem{diamos2010ocelot}
G.~Diamos, A.~Kerr, S.~Yalamanchili, and N.~Clark, ``Ocelot: a dynamic
  optimization framework for bulk-synchronous applications in heterogeneous
  systems,'' in \emph{2010 19th International Conference on Parallel
  Architectures and Compilation Techniques (PACT)}.\hskip 1em plus 0.5em minus
  0.4em\relax IEEE, 2010, pp. 353--364.

\bibitem{han2021cox}
R.~Han, J.~Lee, J.~Sim, and H.~Kim, ``{{COX}}: {{CUDA}} on {{X86}} by exposing
  warp-level functions to {{CPUs}},'' \emph{arXiv preprint arXiv:2112.10034},
  2021.

\bibitem{moses2017should}
W.~S. Moses, ``How should compilers represent fork-join parallelism?'' Master's
  thesis, Massachusetts Institute of Technology, 2017.

\bibitem{schardl2019tapir}
T.~B. Schardl, W.~S. Moses, and C.~E. Leiserson, ``Tapir: Embedding recursive
  fork-join parallelism into llvm’s intermediate representation,'' \emph{ACM
  Transactions on Parallel Computing (TOPC)}, vol.~6, no.~4, pp. 1--33, 2019.

\bibitem{kotsifakou2018hpvm}
M.~Kotsifakou, P.~Srivastava, M.~D. Sinclair, R.~Komuravelli, V.~Adve, and
  S.~Adve, ``{{HPVM}}: Heterogeneous parallel virtual machine,'' in
  \emph{Proceedings of the 23rd ACM SIGPLAN Symposium on Principles and
  Practice of Parallel Programming (PPoPP)}, 2018, pp. 68--80.

\bibitem{stelle2017openmpir}
G.~Stelle, W.~S. Moses, S.~L. Olivier, and P.~McCormick, ``Openmpir:
  Implementing openmp tasks with tapir,'' in \emph{Proceedings of the Fourth
  Workshop on the LLVM Compiler Infrastructure in HPC}, 2017, pp. 1--12.

\bibitem{DBLP:conf/lcpc/DoerfertF18}
\BIBentryALTinterwordspacing
J.~Doerfert and H.~Finkel, ``Compiler optimizations for parallel programs,'' in
  \emph{Languages and Compilers for Parallel Computing - 31st International
  Workshop, {LCPC} 2018, Salt Lake City, UT, USA, October 9-11, 2018, Revised
  Selected Papers}, ser. Lecture Notes in Computer Science, M.~W. Hall and
  H.~Sundar, Eds., vol. 11882.\hskip 1em plus 0.5em minus 0.4em\relax Springer,
  2018, pp. 112--119. [Online]. Available:
  \url{https://doi.org/10.1007/978-3-030-34627-0\_9}
\BIBentrySTDinterwordspacing

\bibitem{DBLP:conf/iwomp/DoerfertDF19}
\BIBentryALTinterwordspacing
J.~Doerfert, J.~M.~M. Diaz, and H.~Finkel, ``The tregion interface and compiler
  optimizations for openmp target regions,'' in \emph{OpenMP: Conquering the
  Full Hardware Spectrum - 15th International Workshop on OpenMP, {IWOMP} 2019,
  Auckland, New Zealand, September 11-13, 2019, Proceedings}, ser. Lecture
  Notes in Computer Science, X.~Fan, B.~R. de~Supinski, O.~Sinnen, and
  N.~Giacaman, Eds., vol. 11718.\hskip 1em plus 0.5em minus 0.4em\relax
  Springer, 2019, pp. 153--167. [Online]. Available:
  \url{https://doi.org/10.1007/978-3-030-28596-8\_11}
\BIBentrySTDinterwordspacing

\bibitem{mlir}
C.~Lattner, M.~Amini, U.~Bondhugula, A.~Cohen, A.~Davis, J.~Pienaar, R.~Riddle,
  T.~Shpeisman, N.~Vasilache, and O.~Zinenko, ``{{MLIR}}: Scaling compiler
  infrastructure for domain specific computation,'' in \emph{2021 IEEE/ACM
  International Symposium on Code Generation and Optimization (CGO)}, 2021, pp.
  2--14.

\bibitem{llvm}
C.~Lattner and V.~Adve, ``{{LLVM}}: a compilation framework for lifelong
  program analysis \& transformation,'' in \emph{International Symposium on
  Code Generation and Optimization, 2004. CGO 2004.}, 2004, pp. 75--86.

\bibitem{mlir_gpu}
\BIBentryALTinterwordspacing
T.~Gysi, C.~M\"{u}ller, O.~Zinenko, S.~Herhut, E.~Davis, T.~Wicky, O.~Fuhrer,
  T.~Hoefler, and T.~Grosser, ``Domain-specific multi-level ir rewriting for
  gpu: The open earth compiler for gpu-accelerated climate simulation,''
  \emph{ACM Trans. Archit. Code Optim.}, vol.~18, no.~4, sep 2021. [Online].
  Available: \url{https://doi.org/10.1145/3469030}
\BIBentrySTDinterwordspacing

\bibitem{polygeist}
W.~S. Moses, L.~Chelini, R.~Zhao, and O.~Zinenko, ``Polygeist: Raising {{C}} to
  polyhedral {{MLIR}},'' in \emph{2021 30th International Conference on
  Parallel Architectures and Compilation Techniques (PACT)}, 2021, pp. 45--59.

\bibitem{rodinia}
S.~Che, M.~Boyer, J.~Meng, D.~Tarjan, J.~W. Sheaffer, S.-H. Lee, and
  K.~Skadron, ``Rodinia: A benchmark suite for heterogeneous computing,'' in
  \emph{2009 IEEE international symposium on workload characterization
  (IISWC)}.\hskip 1em plus 0.5em minus 0.4em\relax Ieee, 2009, pp. 44--54.

\bibitem{he2016deep}
K.~He, X.~Zhang, S.~Ren, and J.~Sun, ``Deep residual learning for image
  recognition,'' in \emph{Proceedings of the IEEE conference on computer vision
  and pattern recognition}, 2016, pp. 770--778.

\bibitem{DBLP:conf/sc/TianSSLGCMSRDZR17}
\BIBentryALTinterwordspacing
X.~Tian, H.~Saito, E.~Su, J.~Lin, S.~Guggilla, D.~Caballero, M.~Masten,
  A.~Savonichev, M.~Rice, E.~Demikhovsky, A.~Zaks, G.~Rapaport, A.~Gaba,
  V.~Porpodas, and E.~N. Garcia, ``{LLVM} compiler implementation for explicit
  parallelization and {SIMD} vectorization,'' in \emph{Proceedings of the
  Fourth Workshop on the {LLVM} Compiler Infrastructure in HPC, LLVM-HPC@SC
  2017, Denver, CO, USA, November 13, 2017}.\hskip 1em plus 0.5em minus
  0.4em\relax {ACM}, 2017, pp. 4:1--4:11. [Online]. Available:
  \url{https://doi.org/10.1145/3148173.3148191}
\BIBentrySTDinterwordspacing

\bibitem{ssa}
\BIBentryALTinterwordspacing
R.~Cytron, J.~Ferrante, B.~K. Rosen, M.~N. Wegman, and F.~K. Zadeck, ``An
  efficient method of computing static single assignment form,'' in
  \emph{Proceedings of the 16th ACM SIGPLAN-SIGACT Symposium on Principles of
  Programming Languages}, ser. POPL '89.\hskip 1em plus 0.5em minus 0.4em\relax
  New York, NY, USA: Association for Computing Machinery, 1989, p. 25–35.
  [Online]. Available: \url{https://doi.org/10.1145/75277.75280}
\BIBentrySTDinterwordspacing

\bibitem{polyhedral}
P.~Feautrier and C.~Lengauer, ``Polyhedron model,'' \emph{Encyclopedia of
  parallel computing}, pp. 1581--1592, 2011.

\bibitem{moses2021reverse}
W.~S. Moses, V.~Churavy, L.~Paehler, J.~H{\"u}ckelheim, S.~H.~K. Narayanan,
  M.~Schanen, and J.~Doerfert, ``Reverse-mode automatic differentiation and
  optimization of gpu kernels via enzyme,'' in \emph{Proceedings of the
  International Conference for High Performance Computing, Networking, Storage
  and Analysis}, 2021, pp. 1--16.

\bibitem{harris2007optimizing}
M.~Harris \emph{et~al.}, ``Optimizing parallel reduction in cuda,''
  \emph{Nvidia developer technology}, vol.~2, no.~4, p.~70, 2007.

\bibitem{doerfert2018compiler}
J.~Doerfert and H.~Finkel, ``Compiler optimizations for openmp,'' in
  \emph{International Workshop on OpenMP}.\hskip 1em plus 0.5em minus
  0.4em\relax Springer, 2018, pp. 113--127.

\bibitem{ppcg_spatial}
\BIBentryALTinterwordspacing
O.~Zinenko, S.~Verdoolaege, C.~Reddy, J.~Shirako, T.~Grosser, V.~Sarkar, and
  A.~Cohen, ``Modeling the conflicting demands of parallelism and
  temporal/spatial locality in affine scheduling,'' in \emph{Proceedings of the
  27th International Conference on Compiler Construction}, ser. CC 2018.\hskip
  1em plus 0.5em minus 0.4em\relax New York, NY, USA: Association for Computing
  Machinery, 2018, p. 3–13. [Online]. Available:
  \url{https://doi.org/10.1145/3178372.3179507}
\BIBentrySTDinterwordspacing

\bibitem{vasilache2012joint}
N.~Vasilache, B.~Meister, M.~Baskaran, and R.~Lethin, ``Joint scheduling and
  layout optimization to enable multi-level vectorization,'' \emph{IMPACT,
  Paris, France}, 2012.

\bibitem{sakamoto2012grand}
K.~Sakamoto and T.~Furumoto, ``Grand central dispatch,'' in \emph{Pro
  Multithreading and Memory Management for iOS and OS X}.\hskip 1em plus 0.5em
  minus 0.4em\relax Springer, 2012, pp. 139--145.

\bibitem{fujitsu_SSLII}
``Fujitsu {SSL} {II} {User's} {Guide} ({Scientific} subroutine library),''
  Fujitsu, Japan.

\bibitem{horovod}
A.~Sergeev and M.~Del~Balso, ``Horovod: fast and easy distributed deep learning
  in tensorflow,'' \emph{arXiv preprint arXiv:1802.05799}, 2018.

\bibitem{benchmarker}
A.~Drozd, ``Benchmarker,'' Online GitHub repository:
  \url{https://github.com/undertherain/benchmarker/}, commit {\tt
  e1f22da320b0c7384cbd2f4df50255c7c2fa6b9d}, 2021.

\bibitem{jaaskelainen2015pocl}
P.~J{\"a}{\"a}skel{\"a}inen, C.~S. de~La~Lama, E.~Schnetter, K.~Raiskila,
  J.~Takala, and H.~Berg, ``pocl: A performance-portable {{OpenCL}}
  implementation,'' \emph{International Journal of Parallel Programming},
  vol.~43, no.~5, pp. 752--785, 2015.

\bibitem{stratton2010efficient}
J.~A. Stratton, V.~Grover, J.~Marathe, B.~Aarts, M.~Murphy, Z.~Hu, and W.-m.~W.
  Hwu, ``Efficient compilation of fine-grained {{SPMD}}-threaded programs for
  multicore {{CPUs}},'' in \emph{Proceedings of the 8th annual IEEE/ACM
  international symposium on Code generation and optimization}, 2010, pp.
  111--119.

\bibitem{karrenberg2012improving}
R.~Karrenberg and S.~Hack, ``Improving performance of {{OpenCL}} on {{CPUs}},''
  in \emph{International Conference on Compiler Construction}.\hskip 1em plus
  0.5em minus 0.4em\relax Springer, 2012, pp. 1--20.

\bibitem{DBLP:conf/cc/MollDH16}
\BIBentryALTinterwordspacing
S.~Moll, J.~Doerfert, and S.~Hack, ``Input space splitting for opencl,'' in
  \emph{Proceedings of the 25th International Conference on Compiler
  Construction, {CC} 2016, Barcelona, Spain, March 12-18, 2016}, A.~Zaks and
  M.~V. Hermenegildo, Eds.\hskip 1em plus 0.5em minus 0.4em\relax {ACM}, 2016,
  pp. 251--260. [Online]. Available:
  \url{https://doi.org/10.1145/2892208.2892217}
\BIBentrySTDinterwordspacing

\bibitem{DBLP:conf/icppw/PatelTDC21}
\BIBentryALTinterwordspacing
A.~Patel, S.~Tian, J.~Doerfert, and B.~M. Chapman, ``A virtual {GPU} as
  developer-friendly openmp offload target,'' in \emph{{ICPP} Workshops 2021:
  50th International Conference on Parallel Processing, Virtual Event / Lemont
  (near Chicago), IL, USA, August 9-12, 2021}, F.~Silla and O.~Marques,
  Eds.\hskip 1em plus 0.5em minus 0.4em\relax {ACM}, 2021, pp. 24:1--24:7.
  [Online]. Available: \url{https://doi.org/10.1145/3458744.3473356}
\BIBentrySTDinterwordspacing

\bibitem{pharr2012ispc}
M.~Pharr and W.~R. Mark, ``ispc: A {{SPMD}} compiler for high-performance
  {{CPU}} programming,'' in \emph{2012 Innovative Parallel Computing
  (InPar)}.\hskip 1em plus 0.5em minus 0.4em\relax IEEE, 2012, pp. 1--13.

\bibitem{beckingsale2019performance}
D.~Beckingsale, R.~Hornung, T.~Scogland, and A.~Vargas, ``Performance portable
  c++ programming with raja,'' in \emph{Proceedings of the 24th Symposium on
  Principles and Practice of Parallel Programming}, 2019, pp. 455--456.

\bibitem{edwards2014kokkos}
H.~C. Edwards, C.~R. Trott, and D.~Sunderland, ``Kokkos: Enabling manycore
  performance portability through polymorphic memory access patterns,''
  \emph{Journal of parallel and distributed computing}, vol.~74, no.~12, pp.
  3202--3216, 2014.

\bibitem{hong2010mapcg}
C.~Hong, D.~Chen, W.~Chen, W.~Zheng, and H.~Lin, ``Mapcg: writing parallel
  program portable between cpu and gpu,'' in \emph{Proceedings of the 19th
  international conference on Parallel architectures and compilation
  techniques}, 2010, pp. 217--226.

\bibitem{10.1145/2627373.2627387}
\BIBentryALTinterwordspacing
A.~Kl\"{o}ckner, ``Loo.py: Transformation-based code generation for {{GPUs}}
  and {{CPUs}},'' in \emph{Proceedings of ACM SIGPLAN International Workshop on
  Libraries, Languages, and Compilers for Array Programming (ARRAY'14)}.\hskip
  1em plus 0.5em minus 0.4em\relax New York, NY, USA: Association for Computing
  Machinery, 2014, p. 82–87. [Online]. Available:
  \url{https://doi.org/10.1145/2627373.2627387}
\BIBentrySTDinterwordspacing

\bibitem{valentin_churavy_2022_6324344}
\BIBentryALTinterwordspacing
V.~Churavy, D.~Aluthge, L.~C. Wilcox, S.~Byrne, M.~Waruszewski, A.~Ramadhan,
  Meredith, S.~Schaub, J.~Schloss, J.~Samaroo, J.~Bolewski, C.~Kawczynski,
  J.~E. Kozdon, J.~Liu, O.~Schulz, Oscar, P.~Haraldsson, T.~Arakaki, and
  T.~Besard, ``Juliagpu/kernelabstractions.jl: v0.8.0,'' Mar. 2022. [Online].
  Available: \url{https://doi.org/10.5281/zenodo.6324344}
\BIBentrySTDinterwordspacing

\bibitem{frigo1998implementation}
M.~Frigo, C.~E. Leiserson, and K.~H. Randall, ``The implementation of the
  cilk-5 multithreaded language,'' in \emph{Proceedings of the ACM SIGPLAN 1998
  conference on Programming language design and implementation}, 1998, pp.
  212--223.

\bibitem{openmpir}
\BIBentryALTinterwordspacing
G.~Stelle, W.~S. Moses, S.~L. Olivier, and P.~McCormick, ``{OpenMPIR}:
  Implementing openmp tasks with tapir,'' in \emph{Proceedings of the Fourth
  Workshop on the LLVM Compiler Infrastructure in HPC}.\hskip 1em plus 0.5em
  minus 0.4em\relax New York, NY, USA: ACM, 2017, pp. 3:1--3:12. [Online].
  Available: \url{http://doi.acm.org/10.1145/3148173.3148186}
\BIBentrySTDinterwordspacing

\bibitem{schmitz2021ppir}
A.~Schmitz, J.~Miller, L.~Tr{\"u}mper, and M.~S. M{\"u}ller, ``Ppir: Parallel
  pattern intermediate representation,'' in \emph{2021 IEEE/ACM International
  Workshop on Hierarchical Parallelism for Exascale Computing (HiPar)}.\hskip
  1em plus 0.5em minus 0.4em\relax IEEE, 2021, pp. 30--40.

\bibitem{stuijk2006sdf}
S.~Stuijk, M.~Geilen, and T.~Basten, ``Sdf\^{} 3: Sdf for free,'' in
  \emph{Sixth International Conference on Application of Concurrency to System
  Design (ACSD'06)}.\hskip 1em plus 0.5em minus 0.4em\relax IEEE, 2006, pp.
  276--278.

\bibitem{moon1999evaluation}
S.~Moon and M.~W. Hall, ``Evaluation of predicated array data-flow analysis for
  automatic parallelization,'' \emph{ACM SIGPLAN Notices}, vol.~34, no.~8, pp.
  84--95, 1999.

\bibitem{oancea2012logical}
C.~E. Oancea and L.~Rauchwerger, ``Logical inference techniques for loop
  parallelization,'' in \emph{Proceedings of the 33rd ACM SIGPLAN conference on
  Programming Language Design and Implementation}, 2012, pp. 509--520.

\bibitem{openmpopt}
{{LLVM Contributors}}, ``{{OpenMP}}-aware optimizations,'' Online:
  \url{https://openmp.llvm.org/optimizations/OpenMPOpt.html}.

\bibitem{10.1145/268946.268974}
\BIBentryALTinterwordspacing
A.~Aiken and D.~Gay, ``Barrier inference,'' in \emph{Proceedings of the 25th
  ACM SIGPLAN-SIGACT Symposium on Principles of Programming Languages}, ser.
  POPL '98.\hskip 1em plus 0.5em minus 0.4em\relax New York, NY, USA:
  Association for Computing Machinery, 1998, p. 342–354. [Online]. Available:
  \url{https://doi.org/10.1145/268946.268974}
\BIBentrySTDinterwordspacing

\bibitem{sorensen2021specifying}
T.~Sorensen, L.~F. Salvador, H.~Raval, H.~Evrard, J.~Wickerson, M.~Martonosi,
  and A.~F. Donaldson, ``Specifying and testing gpu workgroup progress
  models,'' \emph{Proceedings of the ACM on Programming Languages}, vol.~5, no.
  OOPSLA, pp. 1--30, 2021.

\bibitem{sorensen2016portable}
T.~Sorensen, A.~F. Donaldson, M.~Batty, G.~Gopalakrishnan, and
  Z.~Rakamari{\'c}, ``Portable inter-workgroup barrier synchronisation for
  {{GPU}}s,'' in \emph{Proceedings of the 2016 ACM SIGPLAN International
  Conference on Object-Oriented Programming, Systems, Languages, and
  Applications}, 2016, pp. 39--58.

\bibitem{sura2005compiler}
Z.~Sura, X.~Fang, C.-L. Wong, S.~P. Midkiff, J.~Lee, and D.~Padua, ``Compiler
  techniques for high performance sequentially consistent java programs,'' in
  \emph{Proceedings of the tenth ACM SIGPLAN symposium on Principles and
  practice of parallel programming}, 2005, pp. 2--13.

\end{thebibliography}

\end{document}